\documentclass[a4paper,11pt]{article}
\usepackage{TCLreport}
\usepackage{TCLtitlepage}
\usepackage{svg}
\usepackage{subcaption}
\usepackage{graphicx}

\title{Cell Sensing: Traffic detection}
\author{Saúl Jesús Fenollosa Arguedas}

\semester{Spring 2025}
\advisor{Andreas Burg}
\coadvisorA{Sitian Li}
\coadvisorB{}
\coadvisorC{}
\coadvisorD{}
\coadvisorE{}

\begin{document}
\maketitle
\addtocontents{toc}{\protect\thispagestyle{empty}}
\vbox to \textheight{
  \tableofcontents
  \vfill
}
\setcounter{page}{1}

\section{Introduction}

\subsection{Motivation and Context}
Monitoring road traffic in a non-intrusive, scalable, and cost-efficient way remains a major challenge for smart infrastructure. Traditional solutions often rely on physical sensors embedded in the road, cameras with line-of-sight requirements, or dedicated radar units, all of which involve significant deployment and maintenance costs. In contrast, cellular networks are already widely deployed and provide a promising platform for passive sensing applications.

The signals used in fourth-generation Long Term Evolution (LTE) networks, based on Orthogonal Frequency-Division Multiplexing (OFDM), are particularly well-suited for motion sensing due to their fine subcarrier spacing, which allows the detection of small Doppler shifts caused by moving vehicles. These Doppler shifts can be used to estimate vehicle speeds without requiring any active transmission or cooperation from the target.

By leveraging existing LTE infrastructure and hardware, it becomes possible to build a low-cost and fully passive sensing system. The ability to access and process Channel State Information (CSI) from commercial base stations enables the use of everyday communication signals for environmental awareness. This approach aligns with the principles of Integrated Sensing And Communication (ISAC), a growing field that seeks to merge communication and sensing capabilities into a single system to increase spectrum efficiency and system functionality.

Furthermore, the use of multiple receiver channels can significantly enhance the robustness and reliability of signal processing in passive sensing systems. Employing spatially distributed receivers enables differential analysis methods, particularly beneficial for mitigating common impairments such as carrier frequency offset or sampling frequency offset. By comparing phase differences across receiver channels, it becomes possible to isolate motion-induced Doppler shifts more accurately from other phase distortions. Such differential processing not only improves the stability of Doppler estimations in practical scenarios characterized by noise and multipath propagation, but also aligns closely with the broader objectives of accuracy and reliability required for traffic monitoring applications.  

The combination of these factors, including broad infrastructure availability, the inherent motion sensitivity of OFDM, compatibility with ISAC, and improved processing through dual reception, motivates the development of a passive traffic sensing system based on LTE CSI.

\subsection{Problem Statement}

\begin{figure}[htbp]
  \centering
  \includegraphics[width=0.7\linewidth]{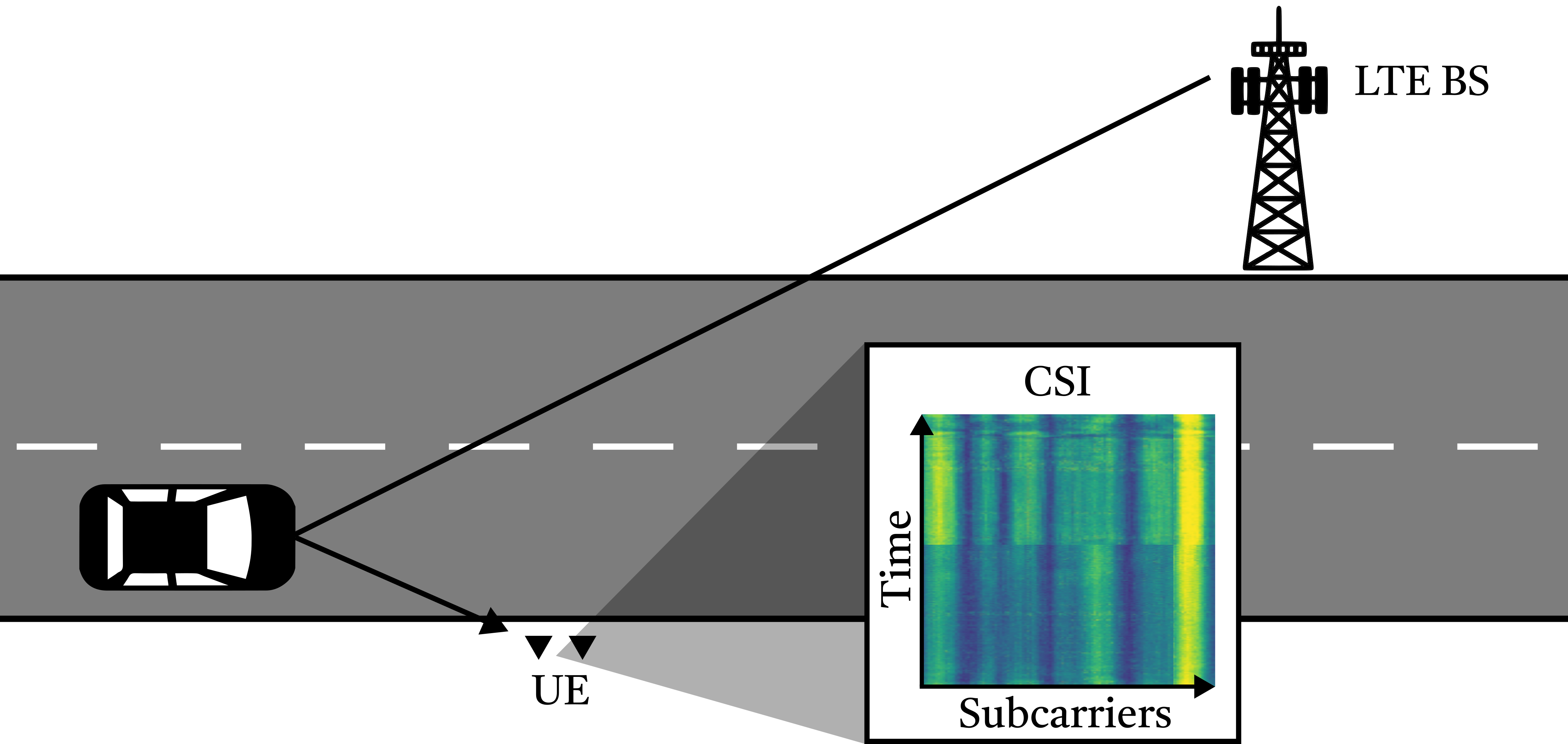}
  \caption{Passive vehicle detection scenario illustration using LTE signals and CSI data.}
  \label{fig:statement}
\end{figure}

This project focuses on the passive detection and classification of traffic conditions, specifically targeting vehicle crossing detection through the induction of Doppler shifts in roadside environments. As illustrated in Figure~\ref{fig:statement}, a vehicle moving through the area between an LTE Base Station (BS) and a receiving User Equipment (UE) induces characteristic patterns in the CSI data, captured across subcarriers and over time. Translating these CSI variations into accurate estimates of vehicle speed involves overcoming practical difficulties. In realistic urban scenarios, CSI signals are typically noisy and influenced by multipath propagation effects. Moreover, signal characteristics vary significantly across measurement sites due to differences in network topology, environmental conditions, and propagation paths. Robust detection and classification thus require algorithms capable of reliably interpreting these subtle and varying CSI patterns, ensuring consistent traffic monitoring across diverse locations.

\subsection{Objectives of the Project}
The main objective of this project is to design a passive sensing system capable of detecting and classifying dynamic events in the environment, such as vehicle motion, through the analysis of CSI acquired from commercial LTE signals. Achieving this objective requires a solid understanding of the radio channel under dynamic conditions, with particular focus on how moving objects influence the temporal and spectral characteristics of the received signal. One key aim is to identify the components of the CSI that are most sensitive to such motion-induced alterations.

To support this, the project proposes the development of a channel model adapted to the sensing application. This model should enable the separation of CSI components associated with transient disturbances, referred to as events, even when the nature or identity of the moving source is unknown. The model is intended to guide both the design of the detection system and the definition of the signal features relevant to motion analysis.

Another key objective is the implementation of a complete system architecture for CSI acquisition, processing, and analysis. This includes both hardware and software elements, structured around a dual-receiver setup to enhance robustness against impairments and to improve sensitivity to motion-related effects. Based on the proposed model, the system should incorporate a dedicated detection algorithm capable of distinguishing meaningful CSI variations from noise and other non-relevant fluctuations. The system is expected to be tested in a controlled indoor environment that emulates real roadside conditions while allowing for repeatable and consistent measurements. This controlled scenario is intended to enable reliable statistical evaluation of performance metrics such as detection rate and speed classification accuracy. Once validated, the system should also be evaluated in real-world outdoor settings, including pedestrian motion in a parking area and vehicle traffic along a roadside, to assess its applicability under operational conditions.

Finally, the project aims to perform a critical evaluation of the results obtained. This includes analyzing the system’s performance across scenarios, identifying its limitations, and extracting insights that may inform future research on CSI-based passive sensing for traffic monitoring.

\subsection{Structure of the report}
This report is structured into 9 sections:

\begin{itemize}
    \item \textbf{Section 2: Background.}  
    Provides the theoretical context for the project, including fundamentals of cellular sensing, an overview of existing traffic detection methods, and a summary of related work and benchmarks. It concludes by outlining the main challenges addressed in this study.

    \item \textbf{Section 3: Theoretical framework.}  
    Presents the radio channel model used in the project, with particular attention to the separation of static and dynamic components in the received CSI. It includes the mathematical formulation that supports the design of the detection method.

    \item \textbf{Section 4: System architecture and design.}  
    Describes the structure of the implemented system, including both hardware and software components. The focus is placed on the dual-receiver setup and the processing pipeline developed for signal filtering and Doppler-based detection.

    \item \textbf{Section 5: Experimental methodology.}  
    Details the measurement scenarios used to evaluate the system. It includes controlled indoor setups and real-world outdoor environments, as well as the procedures followed for ground-truth acquisition and performance evaluation.

    \item \textbf{Section 6: Results.}  
    Presents the outcomes of the experiments in both controlled and outdoor scenarios. The section includes metrics such as detection rate and classification accuracy, supported by statistical analysis.

    \item \textbf{Section 7: Discussion.}  
    Analyzes the system’s performance in light of the results. Particular attention is given to limitations, environmental influences, and the impact of low-speed motion and multipath conditions.

    \item \textbf{Section 8: Future work.}  
    Outlines possible directions for improvement, including the integration of angle-of-arrival estimation, the use of AI techniques, and real-time deployment strategies.

    \item \textbf{Section 9: Conclusion.}  
    Summarizes the main contributions of the project and reflects on its relevance within the context of passive traffic sensing using cellular infrastructure.
\end{itemize}

\newpage
\section{Background}

\subsection{Fundamentals of Wireless Sensing}

Wireless sensing broadly refers to the capability of extracting information about an environment or targets within it by analyzing the propagation of radio frequency (RF) signals \cite{liu2019wireless}. This paradigm has evolved significantly, moving beyond dedicated radar systems to leverage ubiquitous wireless communication infrastructures. Modern cellular networks, characterized by their extensive deployment, sophisticated signal processing capabilities, and standardized protocols, present a compelling platform for opportunistic sensing. Specifically, technologies such as LTE and its successors employ OFDM, a robust modulation scheme that divides the signal bandwidth into numerous narrow subcarriers \cite{sturm2022efficient}. This inherent multi-carrier structure, combined with advanced antenna configurations like Multiple-Input Multiple-Output (MIMO) and beamforming, provides a rich set of channel parameters that are highly sensitive to environmental dynamics \cite{haque2023beamsense}.

At the core of cellular sensing lies the CSI, a fine-grained measurement describing the complex channel matrix between transmitting and receiving antennas across individual subcarriers \cite{halperin2011tool}. Unlike coarser metrics such as Received Signal Strength Indicator (RSSI), CSI encapsulates both amplitude attenuation and phase shifts for each subcarrier, providing a detailed fingerprint of the propagation path. These characteristics are directly influenced by phenomena such as reflection, diffraction, and scattering caused by objects in the environment. Consequently, temporal variations in CSI, particularly in its phase component, can be precisely correlated with the Doppler frequency shifts induced by the relative motion of objects, enabling the detection and characterization of dynamic events \cite{qian2017widance}.

\begin{figure}
    \centering
    \begin{subfigure}[t]{0.25\textwidth}
        \centering
        \includegraphics[width=\textwidth]{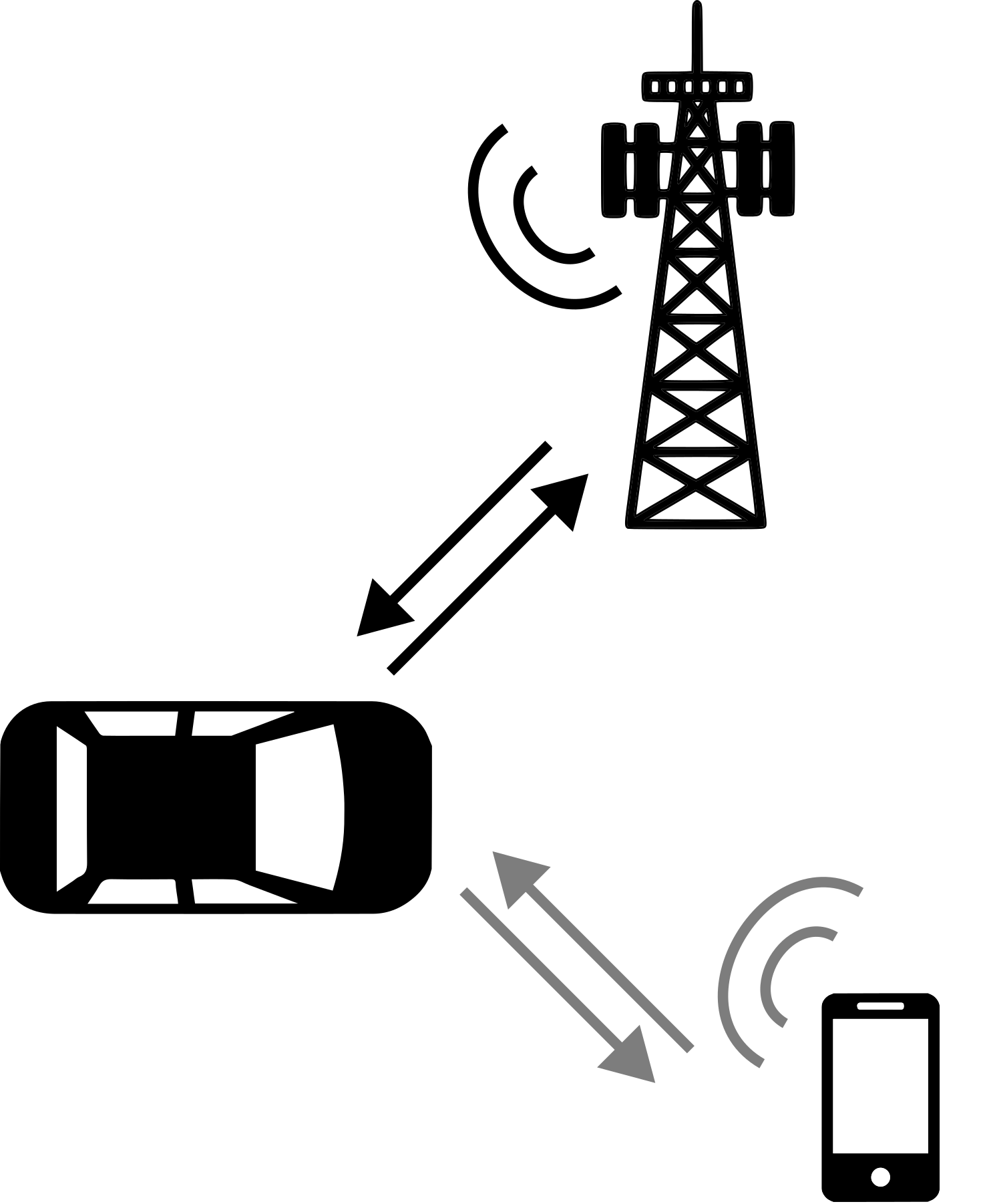}
        \caption{}
    \end{subfigure}
    \hfill
    \begin{subfigure}[t]{0.25\textwidth}
        \centering
        \includegraphics[width=\textwidth]{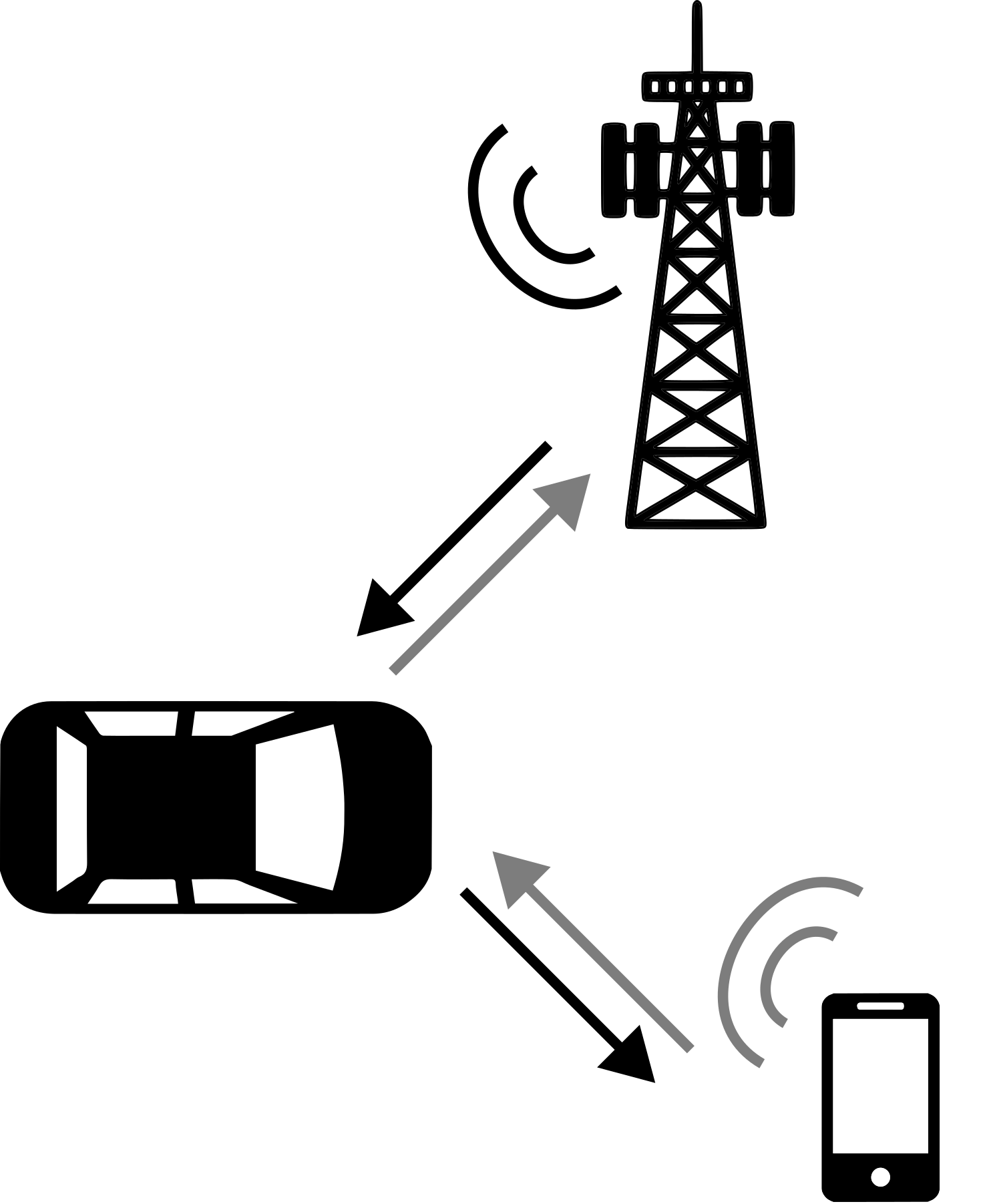}
        \caption{}
    \end{subfigure}
    \hfill
    \begin{subfigure}[t]{0.25\textwidth}
        \centering
        \includegraphics[width=\textwidth]{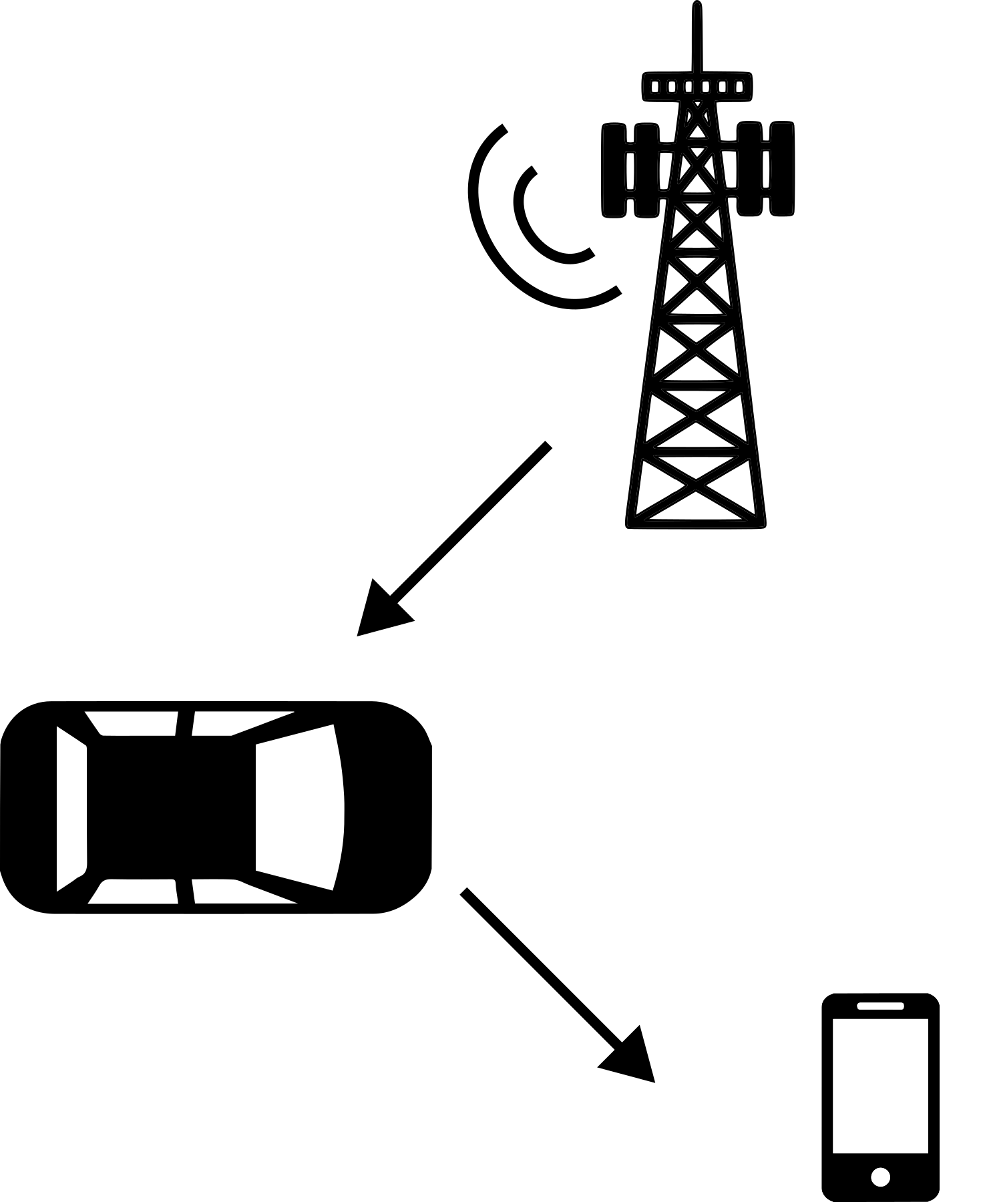}
        \caption{}
    \end{subfigure}
    \caption{Sensing configurations: (a) Monostatic active sensing, (b) Bistatic active sensing, and (c) Bistatic passive sensing.}
    \label{fig:sensing-configs}
\end{figure}

Cellular sensing systems can be broadly categorized based on their operational modes and architectural configurations. Active sensing systems transmit dedicated probing signals and analyze their reflections, akin to traditional radar. In contrast, passive sensing systems capitalize on existing, non-cooperative ambient RF signals (e.g., broadcasting, cellular downlink transmissions) as their illumination source \cite{colone2024passive}. This passive approach offers significant advantages in terms of covertness, spectrum efficiency, and cost-effectiveness, as it obviates the need for dedicated transmitters. Architecturally, sensing systems can be classified as monostatic, where the transmitter and receiver are co-located, or bistatic/multistatic, where they are spatially separated \cite{jiao2025bistatic}. Figure~\ref{fig:sensing-configs} provides a visual summary of these operational and architectural configurations. Passive cellular sensing often inherently operates in a bistatic configuration, as the cellular BS acts as the illuminator and a distinct UE functions as the receiver, analyzing the scattered or reflected signals from targets. This setup can offer unique perspectives and advantages in target detection and localization due to diverse scattering angles.

Despite the inherent advantages, extracting reliable sensing information from CSI is fraught with challenges. The raw phase information embedded within CSI is highly susceptible to various hardware-induced impairments and environmental distortions. These include Carrier Frequency Offset (CFO), Sampling Frequency Offset (SFO), phase noise originating from local oscillators, and I/Q imbalance, all of which introduce undesirable, random phase shifts that can obscure the subtle motion-induced Doppler effects \cite{elbir2021rfimpairments}. Furthermore, practical deployments contend with complex multipath environments, ubiquitous interference, and the persistent presence of static clutter (e.g., walls, furniture), whose strong reflections can overwhelm the weak signals from moving targets \cite{Liu2020Survey}. The instability of sensing performance and the difficulty in generalizing CSI-based models across diverse environments and configurations remain significant open challenges for real-world applications \cite{chen2022solving}.

To overcome these limitations and unlock the full potential of CSI for robust sensing, advanced signal processing and calibration techniques are imperative. A particularly effective strategy involves employing multiple receiver antennas and leveraging their spatial diversity. By analyzing the differential CSI between spatially separated antennas, common-mode phase errors stemming from the receiver chain can be effectively suppressed \cite{patras2021exploiting}. This approach, often referred to as the CSI ratio method or phase difference between antennas, significantly enhances the signal-to-noise ratio of the true motion-induced Doppler shifts, leading to more accurate and reliable sensing outcomes even in noisy and dynamic environments \cite{liu2022csiratio}.

\subsection{Wireless Sensing Approaches for Vehicular Traffic Monitoring}

Building upon the fundamental principles of wireless sensing, the capabilities of CSI-based systems extend significantly to the domain of traffic detection and monitoring. The core challenge in this application lies in accurately converting the subtle, motion-induced changes within the wireless channel into interpretable parameters such as target presence, speed, count, or even classification. This translation often relies on signal processing techniques that exploit specific properties of CSI, notably Doppler frequency shifts and variations in amplitude and phase over time and across subcarriers.

Early advancements in CSI-based motion analysis largely originated from human activity recognition (HAR) systems, which serve as a conceptual stepping stone for traffic applications \cite{liu2019wireless,colone2024passive}. These systems demonstrated that human movements induce unique patterns in CSI, which can be modeled and classified. For instance, the CSI-speed model quantifies the correlation between CSI dynamics and the speed of human movement, effectively inferring velocity from phase-based estimations \cite{wang2015carm}. Analyzing Doppler spectrograms derived from CSI provides a powerful visual and analytical tool, transforming temporal CSI variations into frequency components that directly reveal motion characteristics and distinguish different activities \cite{Li2022csipwr}. While focused on human scale, the methodologies for extracting motion parameters from CSI in these HAR systems are directly transferable to vehicular scenarios, requiring adaptation for different scales of motion and environmental contexts.

The transition from human activity to vehicular traffic detection introduces unique considerations due to higher speeds, larger target sizes, and the more complex outdoor propagation environments. Nevertheless, significant progress has been made in leveraging wireless signals, particularly CSI, for vehicle monitoring. A crucial development in this area involves passive sensing systems that utilize ambient cellular signals. For example, LTE-CommSense demonstrates the feasibility of vehicle detection and classification by exploiting the reference signals inherent in LTE downlink transmissions \cite{sardar2019ltecommsense}. This system operates akin to a passive radar in a forward-scattering radar (FSR) mode, extracting CSI from a standard LTE receiver and analyzing its variations to identify and categorize passing vehicles. This approach provides a robust, non-intrusive, and cost-effective solution for roadside traffic monitoring.

Furthermore, the integration of multi-link or multi-antenna setups has proven invaluable for enhancing the accuracy and scope of CSI-based traffic detection. By deploying multiple wireless links or utilizing multi-antenna configurations, systems can achieve broader sensing coverage and improve the robustness of detection against multipath fading and localized obstructions \cite{murakami2018wlancsimonitoring}. These multi-point observations enable more precise inferences, such as counting objects or estimating crowd density. For instance, approaches combining CSI with RSSI from multiple Wi-Fi receivers have shown efficacy in passenger counting, a technique readily adaptable to vehicular traffic counting scenarios \cite{yang2024rssicsi}. Similarly, multi-antenna CSI has been leveraged for robust occupancy estimation and crowd counting, demonstrating the benefits of spatial diversity for collective motion assessment \cite{romo2021multilinkcrowd}. Such advancements underscore the growing potential of wireless communication infrastructures to serve as versatile platforms for detailed traffic analytics.

\subsection{Current Capabilities and Limitations in CSI Traffic Monitoring}

This section evaluates the current state-of-the-art in wireless traffic sensing by reviewing key performance indicators from existing works, thereby establishing benchmarks for detection accuracy, classification, and speed estimation. Concurrently, it highlights persistent challenges and research gaps that impede the widespread deployment and robust operation of such systems in diverse real-world environments.

Significant progress has been made in leveraging CSI for motion sensing, transitioning from human activity recognition to vehicular traffic monitoring. For instance, the CARM system, utilizing Wi-Fi CSI, demonstrated an average accuracy of 96\% for human speed estimation in indoor environments \cite{wang2015carm}. More directly relevant to vehicular applications, the LTE-CommSense system, employing passive LTE CSI, achieved an average accuracy of detection of 0.94 and an average False Alarm Rate (FAR) of 0.05 for vehicle detection. Furthermore, it attained a classification accuracy of 0.96 for vehicle types, showcasing the feasibility of using ambient cellular signals for comprehensive traffic analysis \cite{sardar2019ltecommsense}. In counting applications, the proposed RSSI-assisted CSI-based passenger counting system achieved an average accuracy and F1-score of over 94\%, specifically 94.86\% accuracy and 94.83\% F1-score, for passenger counting on a double-decker bus with up to 20 passengers \cite{yang2024rssicsi}, with crowd counting systems reporting crowd-counting accuracy rate of over 99\% for up to 8 people \cite{romo2021multilinkcrowd}. While these benchmarks illustrate the potential of wireless sensing, they also implicitly reveal the complexities and limitations inherent in real-world deployments.

Despite these advancements, several critical challenges and open research gaps remain. A primary concern for CSI-based sensing systems is their generalization across diverse environments. Studies indicate a substantial drop in performance, for instance, in user identification, accuracies close to 100\% in training setups can decrease by over 30\% when tested in an unseen furniture setting. Similarly, for gesture recognition, baseline models trained in one environment exhibited a 27.2\% accuracy when applied to a different, unseen room due to significant environmental variations \cite{chen2022solving}. This highlights the inherent susceptibility of wireless signals to environmental variations and underscores the necessity for highly robust algorithms capable of maintaining performance consistency across different deployment locations and propagation conditions.

Furthermore, current methodologies often focus solely on speed estimation, with direction detection remaining an underexplored area for vehicular traffic. The inherent limitations of static clutter subtraction methods further complicate accurate motion analysis in dynamic environments. Beyond signal processing, practical issues in deploying and scaling CSI-based sensing systems, including maintaining precise synchronization across distributed receivers in ISAC contexts and addressing privacy concerns, present considerable challenges \cite{jiao2025bistatic,Cerutti2022PrivacyCSI}. Lastly, despite advancements in hardware impairment mitigation, the persistent need for robust phase sanitization techniques underscores a fundamental challenge in extracting reliable motion information from noisy CSI signals \cite{Li2025ComplexBeat}. Addressing these gaps is crucial for the transition of CSI-based traffic monitoring from experimental prototypes to widely deployable, reliable solutions.

\newpage
\section{Theoretical Framework}

\subsection{Single-RX CSI Channel Model}

Modeling the wireless propagation channel in dynamic environments, where the presence and motion of objects cause its characteristics to vary over time, is a fundamental challenge in wireless sensing. A cornerstone for understanding such time-variant linear channels is the seminal work by P.A. Bello \cite{bello1963channel}. Bello's framework introduces the concept of the delay-Doppler spread function, $S(\tau, \nu)$, which characterizes the channel in terms of signal energy distribution across propagation delays ($\tau$) and Doppler frequency shifts ($\nu$). This function provides a powerful representation of how a wireless signal is dispersed in both time (due to multipath propagation) and frequency (due to motion).

The channel's instantaneous time-variant impulse response, $h_j(\tau, t)$, at a receiver antenna $j$, describes the channel's response at time $t$ to an impulse arriving with delay $\tau$. In a scattering environment, this impulse response can be conceptualized as a superposition of contributions from $N$ discrete propagation paths, each originating from a distinct scatterer. For a simplified scenario where each scatterer $n$ contributes a specific complex amplitude $a_{j,n}$, a constant propagation delay $\tau_{j,n}$, and a Doppler frequency shift $\nu_{j,n}$ (induced by its relative velocity with respect to the transmitter-receiver link), the time-variant impulse response for receiver $j$ can be expressed as:
\begin{equation}
h_j(\tau, t) = \sum_{n=1}^{N} a_{j,n} \delta(\tau - \tau_{j,n}) e^{j2\pi\nu_{j,n}t}
\label{eq:impulse_response}
\end{equation}
Here, $\delta(\cdot)$ is the Dirac delta function, indicating that each path contributes at a specific delay.

The $\hat{H}_{j}(t,f)$ represents the channel's instantaneous frequency response. It is obtained by taking the Fourier Transform of the time-variant impulse response $h_j(\tau, t)$ with respect to the delay variable $\tau$. Applying the Fourier Transform to the sum of individual path contributions yields:
\begin{equation}
\mathcal{F}_{\tau} \{ h_j(\tau, t) \} = \mathcal{F}_{\tau} \left\{ \sum_{n=1}^{N} a_{j,n} \delta(\tau - \tau_{j,n}) e^{j2\pi\nu_{j,n}t} \right\}
\end{equation}
Given that the Fourier Transform of $\delta(\tau - \tau_{j,n})$ is $e^{-j2\pi f \tau_{j,n}}$, and treating $t$ as fixed for the instantaneous response, we obtain the channel frequency response:
\begin{equation}
H_{j}(t,f) = \sum_{n=1}^{N} a_{j,n} e^{j2\pi\nu_{j,n}t} e^{-j2\pi f\tau_{j,n}}
\label{eq:channel_freq_response}
\end{equation}

In practical scenarios, the wireless channel typically comprises contributions from both moving and stationary objects. Stationary scatterers are those for which the Doppler frequency shift $\nu_{j,n}$ is zero. The terms corresponding to these zero-Doppler components, where $e^{j2\pi\nu_{j,n}t} = 1$, effectively form a time-invariant frequency response for these specific paths, which can be grouped into a static channel part, denoted as $H_{s,j}(f)$. For clarity and to facilitate subsequent signal processing steps aimed at distinguishing static background clutter from dynamic targets, it is conventional and analytically convenient to separate these components in the channel model.

Once the static part is separated, the summation index is restricted to the $M$ \textit{dynamic} paths only. However, the actually measured CSI, $\hat{H}_{j}(t,f)$, is not solely the true channel response. It is additionally corrupted by various hardware imperfections intrinsic to the receiver's RF front-end. These impairments, which are often time- and frequency-dependent, introduce multiplicative distortions. They include the already mentioned CFO, SFO, phase noise originating from local oscillators, and I/Q imbalance. These combined hardware-induced effects are represented by a multiplicative complex term, $C_j(t,f)$. Considering discrete time and frequency samples, where $t=nT_{\text{sym}}$ (with $n$ being the OFDM symbol index and $T_{\text{sym}}$ the symbol duration) and $f=f_{c}+k\Delta f$ (with $f_c$ being the center frequency and $k$ the subcarrier index within a bandwidth $\Delta f$), the single-receiver CSI model can be rigorously expressed as:
\begin{equation}
\hat{H}_{j}(t,f) = C_j(t,f)\left(H_{s,j}(f) + \sum_{m=1}^{M} a_{j,m}e^{j2\pi\nu_{j,m}t}e^{-j2\pi f\tau_{j,m}} \right)
\label{eq:measured_csi}
\end{equation}
where $m$ now indexes only the $M$ dynamic (i.e., moving) scatterers.

A critical limitation of this single-receiver configuration lies in the multiplicative nature of the $C_j(t,f)$ term. The random and often substantial phase variations introduced by these hardware impairments effectively obscure the much smaller, yet crucial, phase shifts caused by Doppler effects. Consequently, the direct phase information derived from a single antenna system is inherently unreliable for accurate Doppler estimation and robust motion sensing. This fundamental limitation underscores the critical need for receiver diversity, typically implemented through a dual-receiver (dual-RX) setup, to enable effective cancellation of these common-mode phase errors and unlock the precise Doppler information required for advanced sensing applications.

\subsection{Dual-RX Calibration and Doppler-difference Derivation}

The fundamental limitation of single-receiver CSI systems lies in the corruption of motion-induced phase information by receiver-specific hardware impairments. To overcome this, a dual-RX configuration is commonly employed. The core principle of dual-RX calibration relies on the assumption that the hardware impairments, represented by the multiplicative term $C_j(t,f)$, are largely common or highly correlated between the two closely spaced receivers. That is, $C_0(t,f) \approx C_1(t,f) \approx C(t,f)$. This assumption enables the cancellation of these common-mode phase errors through differential processing.

Let the measured CSI at receiver 0 (Rx0) and receiver 1 (Rx1) be denoted as $\hat{H}_{0}(t,f)$ and $\hat{H}_{1}(t,f)$, respectively. Based on the single-RX CSI channel model established in the previous subsection, considering a dominant dynamic path (echo) for simplicity, these can be expressed as:
\begin{align*}
\hat{H}_{0}(t,f) &= C(t,f)\left(H_{s,0}(f) + a_{0}e^{j2\pi\nu_{0}t}e^{-j2\pi f\tau_{0}} \right) \\
\hat{H}_{1}(t,f) &= C(t,f)\left(H_{s,1}(f) + a_{1}e^{j2\pi\nu_{1}t}e^{-j2\pi f\tau_{1}} \right)
\end{align*}
where $H_{s,j}(f)$ denotes the static channel part for receiver $j$, and $a_{j}$, $\nu_{j}$, $\tau_{j}$ correspond to the amplitude, Doppler frequency, and delay of the dominant dynamic echo for receiver $j$, respectively. The hardware impairment term $C(t,f)$ is assumed common to both receivers.

To cancel the common hardware impairments and extract meaningful differential Doppler information, a cross-correlation operation is performed by multiplying the CSI of one receiver with the complex conjugate of the other. This results in a new composite channel response, $\tilde{H}(t,f)$:
\begin{align*}
\tilde{H}(t,f) &= \hat{H}_{1}(t,f)\hat{H}_{0}^{*}(t,f) \\
&= \left[ C(t,f)\left(H_{s,1}(f) + a_{1}e^{j2\pi\nu_{1}t}e^{-j2\pi f\tau_{1}} \right) \right] \left[ C(t,f)\left(H_{s,0}(f) + a_{0}e^{j2\pi\nu_{0}t}e^{-j2\pi f\tau_{0}} \right) \right]^{*} \\
&= |C(t,f)|^2 \left(H_{s,1}(f) + a_{1}e^{j2\pi\nu_{1}t}e^{-j2\pi f\tau_{1}} \right) \left(H_{s,0}^{*}(f) + a_{0}^{*}e^{-j2\pi\nu_{0}t}e^{j2\pi f\tau_{0}} \right)
\end{align*}
Expanding the product of the two terms within the parentheses, we obtain four distinct components:
\begin{align*}
\tilde{H}(t,f) &= |C(t,f)|^2 \Big[ H_{s,1}(f)H_{s,0}^{*}(f) \\
& \quad + H_{s,1}(f)a_{0}^{*}e^{-j2\pi\nu_{0}t}e^{j2\pi f\tau_{0}} \\
& \quad + H_{s,0}^{*}(f)a_{1}e^{j2\pi\nu_{1}t}e^{-j2\pi f\tau_{1}} \\
& \quad + a_{1}a_{0}^{*}e^{j2\pi\nu_{1}t}e^{-j2\pi f\tau_{1}}e^{-j2\pi\nu_{0}t}e^{j2\pi f\tau_{0}} \Big]
\end{align*}
By rearranging the terms in the last component, specifically grouping the exponential terms related to Doppler frequencies and delays, we arrive at the full expression as visually presented:\newline
\resizebox{\textwidth}{!}{$
\tilde{H}(t,f) = |C(t,f)|^2 \left[ 
\underbrace{H_{s,1}H_{s,0}^{*}}_{\text{static}} + 
\underbrace{H_{s,1}a_{0}^{*}e^{-j2\pi\nu_{0}t}e^{j2\pi f\tau_{0}}}_{-\nu_0} + 
\underbrace{H_{s,0}^{*}a_{1}e^{j2\pi\nu_{1}t}e^{-j2\pi f\tau_{1}}}_{+\nu_1} + 
\underbrace{a_{1}a_{0}^{*}e^{j2\pi(\nu_{1}-\nu_{0})t}e^{-j2\pi f(\tau_{1}-\tau_{0})}}_{\Delta\nu}
\right]
$}\vspace{5pt}

This composite CSI, $\tilde{H}(t,f)$, now contains terms that are explicitly related to the difference in Doppler shifts, $\Delta\nu = \nu_1 - \nu_0$, and the difference in delays, $\Delta\tau = \tau_1 - \tau_0$, between the dominant dynamic paths observed by the two receivers. The crucial advantage of this multiplication is that the common hardware impairment term $C(t,f)$ is transformed into its magnitude squared, $|C(t,f)|^2$, which is a real, non-oscillatory term, thereby effectively removing the problematic random phase oscillations that plagued the single-receiver measurement.

In vehicular traffic detection applications, attention is given to scenarios where vehicles move approximately perpendicular to the normal vector of the plane defined by two receiving antennas. This geometry is representative of typical roadside monitoring configurations. In such setups, the \textit{differential dynamic cross-term}, expressed as
\[
a_{1}a_{0}^{*}e^{j2\pi(\nu_{1}-\nu_{0})t}e^{-j2\pi f(\tau_{1}-\tau_{0})},
\]
becomes a key component. The phase derivative of this term with respect to time is proportional to the instantaneous differential Doppler shift
\[
\Delta\nu(t)=\nu_1(t)-\nu_0(t).
\]

To illustrate its relevance, consider a moving reflector with constant velocity \(\vec{v}\) along a linear path. The Doppler frequency at receiver \(j\) due to this reflector is approximated as
\[
\nu_j(t)=-\frac{1}{\lambda}\frac{d\!\left(R_{\text{Tx}\to R}(t)+R_{R\to\text{Rx}_j}(t)\right)}{dt},
\]
where \(R_{\text{Tx}\to R}\) and \(R_{R\to\text{Rx}_j}\) denote the distances from the transmitter to the reflector and from the reflector to receiver \(j\), respectively. When the transmitter is positioned sufficiently far from the receivers, the derivative of \(R_{\text{Tx}\to R}\) contributes equally to \(\nu_1\) and \(\nu_0\) and thus cancels in the differential term. Consequently, the differential Doppler is dominated by
\[
\Delta\nu(t)\approx-\frac{1}{\lambda}\frac{d\!\left(R_{R\to\text{Rx}_1}(t)-R_{R\to\text{Rx}_0}(t)\right)}{dt}.
\]

Let the receiver positions be \((x_0,y_{\text{Rx}})\) and \((x_1,y_{\text{Rx}})\), and let the reflector be located at \((x_R(t),y_R)\). The range to receiver \(j\) is given by
\[
R_{R\to\text{Rx}_j}(t)=\sqrt{(x_R(t)-x_j)^2+(y_R-y_{\text{Rx}})^2}.
\]
Differentiating this with respect to time yields
\[
\frac{dR_{R\to\text{Rx}_j}}{dt}=\frac{v_x\!\left(x_R(t)-x_j\right)}{R_{R\to\text{Rx}_j}(t)},
\]
where \(v_x\) is the reflector's velocity component along the \(x\)-axis. Hence, the differential Doppler becomes
\[
\Delta\nu(t)\propto v_x\left(\frac{x_R(t)-x_0}{R_{R\to\text{Rx}_0}(t)}-\frac{x_R(t)-x_1}{R_{R\to\text{Rx}_1}(t)}\right).
\]

The sign of \(\Delta\nu(t)\) is also informative, because it preserves the direction of motion projected onto the antenna baseline: a positive \(\Delta\nu\) indicates that the reflector is moving from \(\text{Rx0}\) toward \(\text{Rx1}\) (i.e., \(v_x>0\)), whereas a negative value implies the opposite.

\begin{figure}[htbp]
  \centering
  \includegraphics[width=0.5\linewidth]{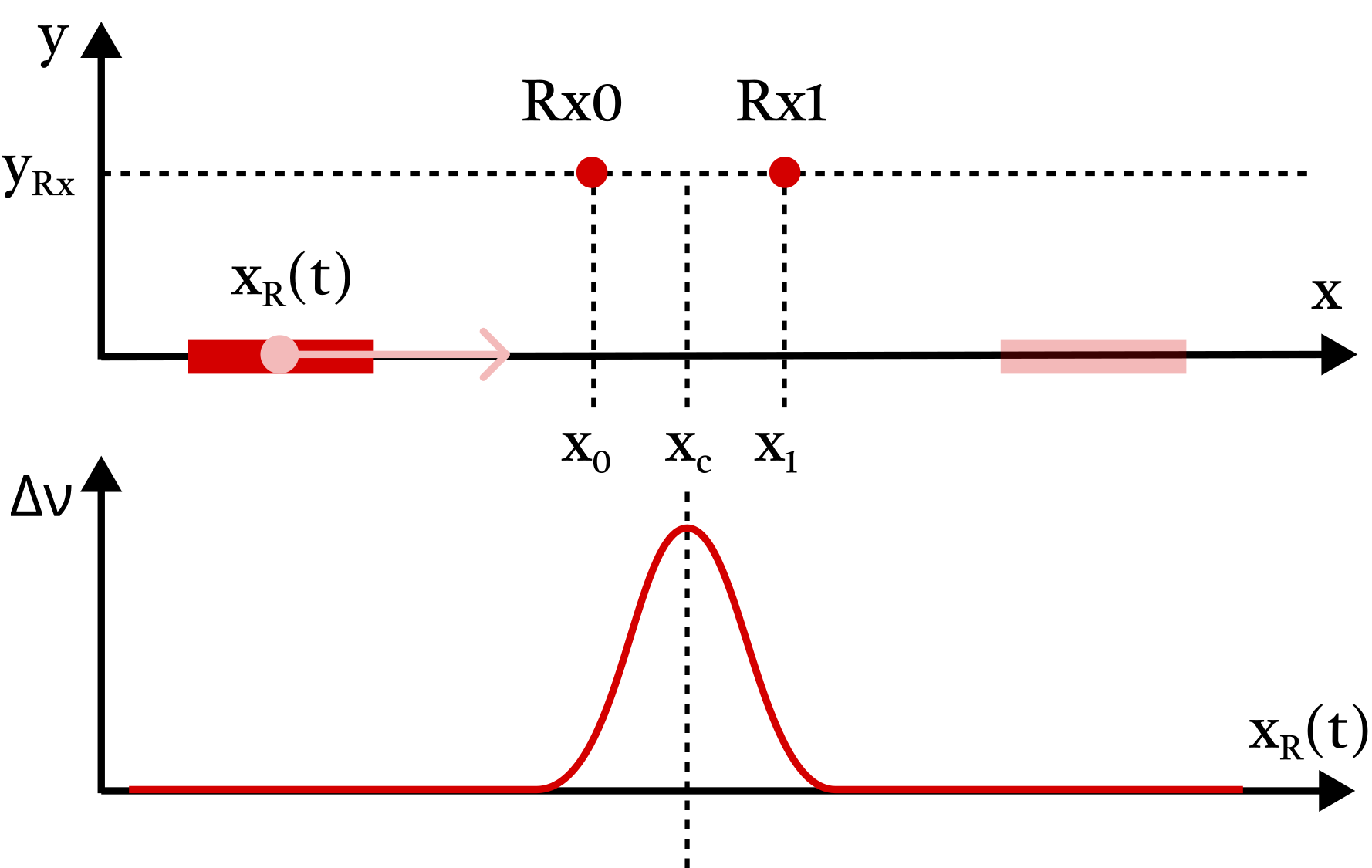}
  \caption{Passive vehicle detection scenario illustration using LTE signals and CSI data.}
  \label{fig:trajectory}
\end{figure}

When the reflector crosses the midpoint \(x_c=(x_0+x_1)/2\) between the receivers, the terms \((x_R(t)-x_0)\) and \((x_R(t)-x_1)\) are equal in magnitude and opposite in sign, and the distances to both receivers are equal. This leads to a maximized absolute value of \(\Delta\nu(t)\) at \(x_R=x_c\), providing a distinct peak in the differential Doppler profile that serves as a robust indicator of a vehicle crossing the antenna baseline. Figure~\ref{fig:trajectory} illustrates the described geometry and the resulting differential-Doppler peak.

Furthermore, under this midpoint condition, where \(R_{R\to\text{Rx}_0}=R_{R\to\text{Rx}_1}=R_m\), the expression simplifies to
\[
\Delta\nu(x_c)\approx-\frac{v_x}{\lambda}\left(\frac{x_c-x_0}{R_m}-\frac{x_c-x_1}{R_m}\right)=-\frac{v_x}{\lambda}\frac{2(x_c-x_0)}{R_m}.
\]
This yields an explicit estimate of the reflector's velocity:
\[
v_x\approx-\Delta\nu(x_c)\frac{\lambda R_m}{2(x_c-x_0)}.
\]

The prominence of the differential dynamic cross-term in the composite frequency response \(\tilde{H}(t,f)\) is essential for reliable detection. The conjugate multiplication between two receivers converts the hardware-related complex coefficient \(C(t,f)\) into a real-valued magnitude square, \(|C(t,f)|^2\), thus eliminating phase errors due to receiver-specific impairments. However, to isolate the dynamic component effectively, the static-static term \(H_{s,1}H_{s,0}^{*}\) and the static-dynamic cross-terms must be suppressed. This is achieved via background subtraction techniques that attenuate static contributions \cite{sardar2019ltecommsense}.

Comparable strategies are employed in applications such as respiration sensing (e.g., \cite{zeng2018farsense}), where the dynamic signals are weak and often obscured by dominant static paths. There, antenna symmetry and close placement enable cancellation of the Line-of-Sight (LOS) path. In contrast, vehicle reflections in traffic monitoring scenarios are typically stronger, making the differential approach particularly suitable. This model proves effective when:
\vspace{-1pt}
\begin{itemize}
  \item Static background subtraction is feasible,
  \vspace{-2pt}
  \item The vehicle echo is strong enough to be discerned, and
  \vspace{-2pt}
  \item The scenario involves a dominant dynamic path.
\end{itemize}
\vspace{-1pt}
By isolating and emphasizing the differential dynamic cross-term, the model offers a robust mechanism for detecting vehicular passage, estimating velocity, and inferring motion direction, which are foundational capabilities in passive traffic monitoring systems.

\newpage
\section{System Architecture \& Data Analysis Methodology}

\subsection{Hardware platform}

\begin{figure}[htbp]
    \centering
    \begin{subfigure}[t]{0.45\textwidth}
        \centering
        \includegraphics[width=0.9\textwidth]{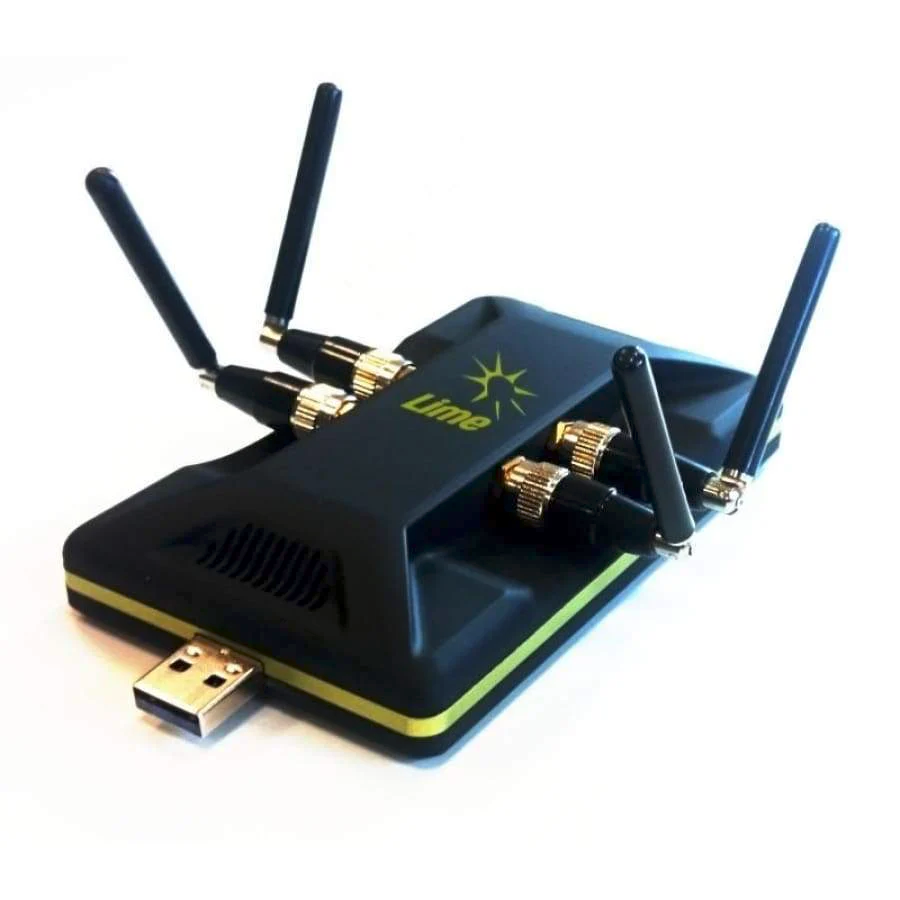}
        \caption{}
        \label{fig:limesdr}
    \end{subfigure}
    \hfill
    \begin{subfigure}[t]{0.45\textwidth}
        \centering
        \includegraphics[width=0.9\textwidth]{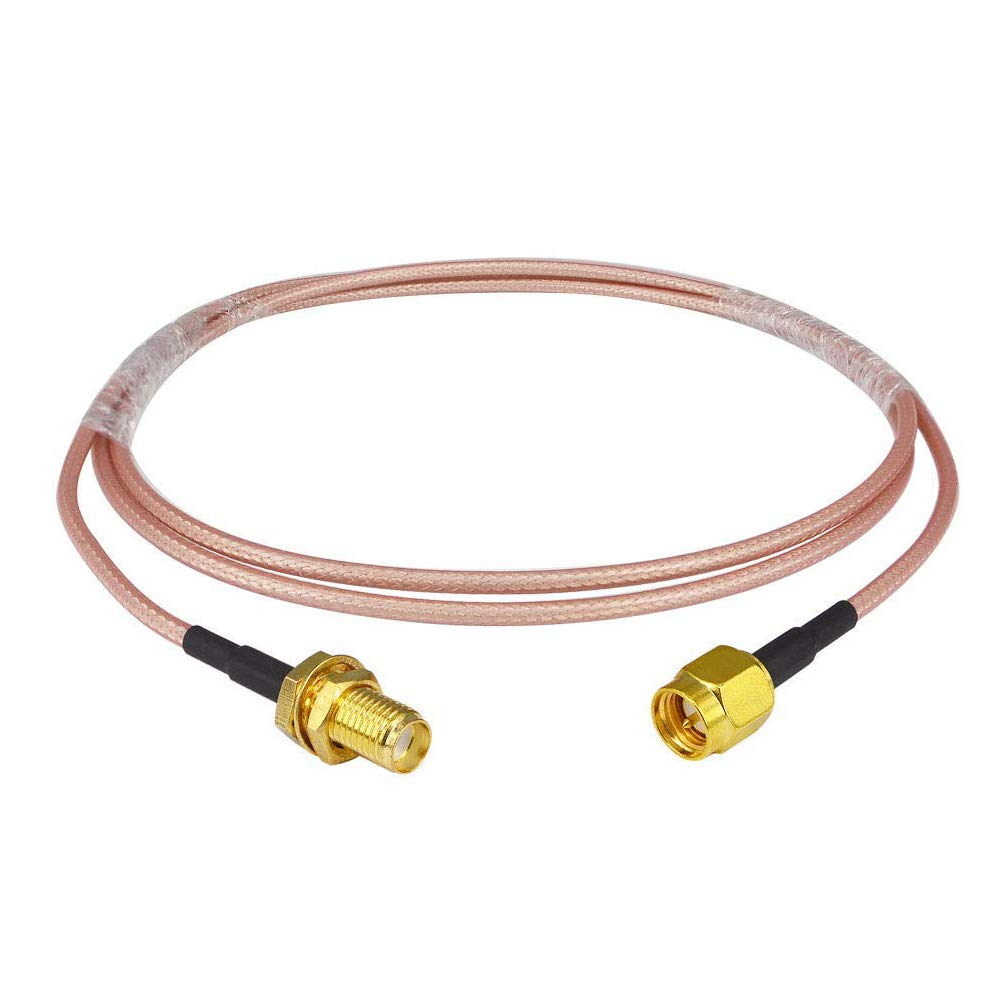}
        \caption{}
        \label{fig:rg316-cables}
    \end{subfigure}
    \caption{Hardware components of the passive sensing system: (a) LimeSDR USB with LMS7002M RF front-end, and (b) RG316 coaxial cables.}
    \label{fig:hardware-components}
\end{figure}

The system's core hardware is the LimeSDR USB board, which incorporates the LMS7002M RF transceiver \cite{attify2025limesdrcase}. As depicted in Figure~\ref{fig:limesdr}, the LimeSDR USB functions as a Software Defined Radio (SDR) platform. The embedded LMS7002M is a field-programmable RF (FPRF) integrated circuit that provides two transmit and two receive chains, enabling a $2 \times 2$ MIMO configuration. This dual-channel capability supports the differential processing techniques discussed in Section 3, facilitating the simultaneous acquisition of CSI from two spatially separated receiver antennas.

The board's frequency range extends from $100 \text{ kHz}$ to $3.8 \text{ GHz}$. This range covers various wireless communication bands, including the LTE Band 1 (at $2.1 \text{ GHz}$) utilized in this project. A programmable bandwidth of up to $61.44 \text{ MHz}$ allows for the capture of multiple OFDM subcarriers, which supports CSI reconstruction and Doppler information extraction. The LMS7002M's sampling rate and dynamic range further contribute to the digitization of received RF signals, influencing quantization noise levels and signal integrity.

For signal reception, the LimeSDR USB utilizes omnidirectional antennas, each providing a typical gain ranging from $2 \text{ to } 3 \text{ dBi}$. Omnidirectional antennas acquire signals uniformly across horizontal directions. This characteristic is relevant in passive sensing scenarios where the direction of arrival of reflections is not predetermined. The specified gain range supports a broad reception pattern.

Two RG316 coaxial cables connect the antennas to the LimeSDR board (Figure~\ref{fig:rg316-cables}). These cables offer flexibility, a small diameter, and compatibility with the operating frequency range. Coaxial cables introduce signal attenuation, a factor in maintaining signal quality, particularly over longer distances. The specified attenuation for RG316 cables is $1.247 \text{ to } 1.903 \text{ dB/m}$ at relevant frequencies \cite{fairview2020rg316}. This characteristic requires consideration in the overall link budget, as it influences received signal strength and the signal-to-noise ratio (SNR) of acquired CSI.

In summary, the hardware platform, consisting of the LimeSDR USB with the LMS7002M transceiver, omnidirectional antennas, and RG316 coaxial cables, supports the passive LTE CSI sensing system. Its multi-channel capability, frequency and bandwidth support, and raw IQ sample acquisition facilitate the dual-RX calibration and Doppler-difference derivation techniques utilized for traffic monitoring.

\subsection{Software platform: srsRAN 4G}

SDR platforms facilitate wireless communications research by enabling protocol stack implementations in software on general-purpose processors. The srsRAN 4G suite is an open-source, full-stack implementation of a 4G LTE network, developed in C++ for Linux-based systems operating on commodity hardware. This suite comprises three primary, interoperable components: srsEPC (Evolved Packet Core), srsENB (Evolved Node B), and srsUE (User Equipment).

The srsUE application serves as the user equipment component, implementing the complete LTE protocol stack for UE-eNodeB interaction. It interfaces with the physical layer using SDR hardware drivers, typically through the SoapySDR library for devices like the LimeSDR, to stream raw In-phase and Quadrature (I/Q) radio samples. This enables a standard computer and SDR device to emulate an LTE user device, allowing access to wireless protocol layers.

The internal architecture of srsUE follows the 3GPP-defined layered protocol stack. The Physical (PHY) layer is responsible for radio signal transmission and reception. This includes channel estimation, which involves analyzing known reference signals (such as CSI-RS or Cell-Specific Reference Signals, CRS) in the downlink signal to characterize the radio channel. This process generates raw CSI data. Higher layers manage communication flow; for example, the Medium Access Control (MAC) layer handles scheduling and error correction, and the Radio Link Control (RLC) and Packet Data Convergence Protocol (PDCP) layers facilitate data transfer to the Internet Protocol (IP) layer.

The Radio Resource Control (RRC) layer manages CSI acquisition by the UE. The PHY layer performs channel measurements, but CSI reporting is controlled by the network. The eNodeB transmits RRCConnectionReconfiguration messages to the UE, containing instructions for CSI reporting configuration, such as periodicity and report type. srsUE's RRC layer processes these instructions and configures its MAC and PHY layers for measurement and reporting. This control mechanism requires eNodeB configuration to enable CSI capture.

For this project, fine-grained CSI data is extracted from the PHY layer's channel estimation. In addition to summarized CSI metrics (e.g., Channel Quality Indicator (CQI), Rank Indicator (RI), Reference Signal Received Power (RSRP), SNR), the system acquires complex-valued CSI matrices per OFDM block and subcarrier. The data writting has been formatted as follows:

\begin{verbatim}
[ESTIMATION]
Timestamp: 1743022991574101
SNR: 3.235169
RSRP: 56.205956
Cell Parameters: center_freq_Hz=21302999040.000000, nof_prb=100, 
cp=normal, symbol_sz=1536, useful_re=1200, offset=0, ofdm_symbols=14
subcarrier_stride: 1, block_stride: 2
[PORT 0]
[RX ANTENNA 0]
OFDM_Block 0: (-12.185438,0.139236), (-12.498656,-1.154930), ...
OFDM_Block 2: ...
[RX ANTENNA 1]
...
[PORT 1]
...
[END ESTIMATION]
\end{verbatim}

This structure provides raw complex coefficients of the channel frequency response, supporting analysis of phase and amplitude variations for Doppler shift estimation. Captured CSI data from multiple experimental runs is organized into a nested dictionary-like structure for processing. This format allows for stacking and manipulation of time-series data, with CSI per port and receiver antenna stored as NumPy arrays:

\begin{verbatim}
{
  'blocks': [
    {
      'timestamp': ...,
      'snr': ...,
      'rsrp': ...,
      'center_freq_hz': ...,
      'nof_prb': ...,
      'symbol_sz': ...,
      'useful_re': ...,
      'offset': ...,
      'ofdm_symbols': ...,
      'subcarrier_stride': ...,
      'block_stride': ...,
      'port_data': {
        (port, rx): {
          'ofdm_block_indices': [...],
          'csi': numpy array with shape (#ofdm_blocks, #subcarriers)
        },
        ...
      }
    },
    ...
  ]
}
\end{verbatim}

This approach enables the use of complex-valued CSI matrices for channel modeling and motion detection in subsequent processing stages.

\subsection{Methodology for Differential Doppler Extraction}

\begin{figure}[htbp]
    \centering
    \includegraphics[width=\textwidth]{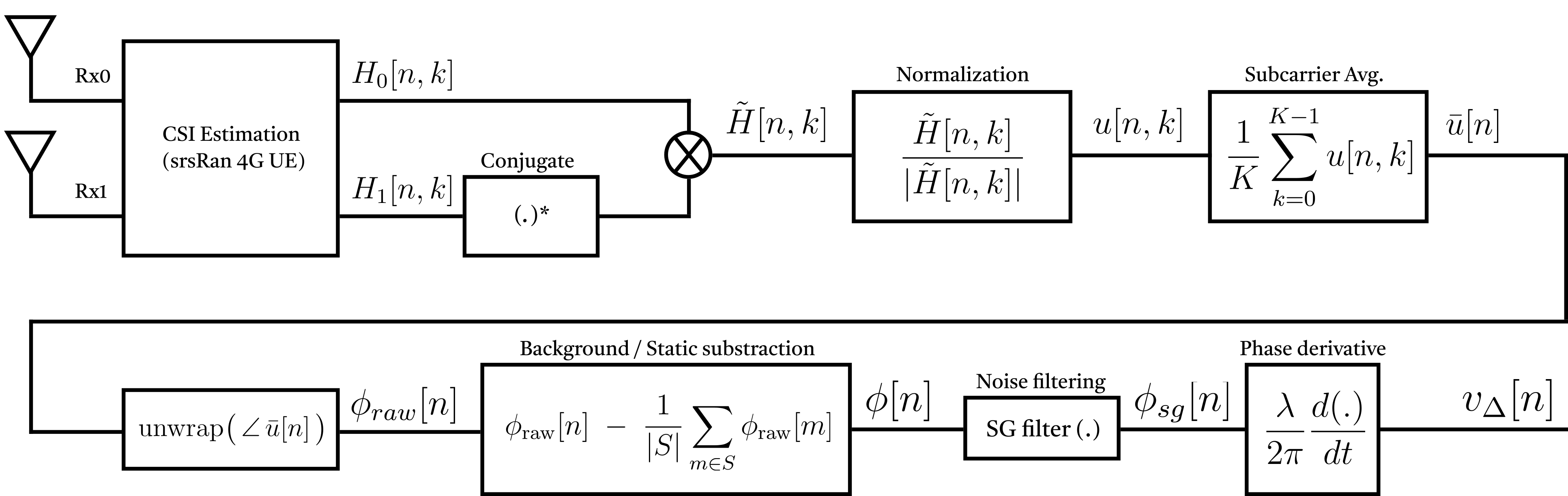}
    \caption{Signal Processing Pipeline for Differential Doppler Extraction.}
    \label{fig:processing_pipeline}
\end{figure}

The recovery of differential Doppler from raw CSI measurements is a process designed to isolate the dynamic changes in the wireless channel from pervasive noise, hardware impairments, and environmental clutter. This section provides a detailed exposition of each stage in the signal processing pipeline, demonstrating how the signal is progressively transformed to yield a differential Doppler estimate. The complete sequence of processing steps is illustrated in Figure~\ref{fig:processing_pipeline}.

The input to this pipeline is the composite signal $\tilde{H}(t,f)$, which is obtained after the dual-RX multiplication of the raw CSI from two receivers. As derived in Section 3.2, this signal incorporates the true channel response, hardware impairments, and additive noise, and can be expressed as:\vspace{10pt}
\newline
\resizebox{\textwidth}{!}{
$\tilde{H}(t,f) = |C(t,f)|^2 \left[ H_{s,1}H_{s,0}^{*} + H_{s,1}a_{0}^{*}e^{j\Phi_{sd0}(t,f)} + H_{s,0}^{*}a_{1}e^{j\Phi_{sd1}(t,f)} + a_{1}a_{0}^{*}e^{j\Phi_{dd}(t,f)} \right] + N_{prod}(t,f)$}
\vspace{0pt} \newline where $|C(t,f)|^2$ represents the common hardware impairment term after cancellation, $H_{s,j}(f)$ is the static channel component for receiver $j$, $a_{j,m}$ are dynamic amplitudes, $\Phi_{sd0}(t,f) = -2\pi\nu_0 t + 2\pi f\tau_0$ and $\Phi_{sd1}(t,f) = 2\pi\nu_1 t - 2\pi f\tau_1$ are phases for static-dynamic cross-terms, and $\Phi_{dd}(t,f) = 2\pi(\nu_1-\nu_0)t - 2\pi f(\tau_1-\tau_0)$ is the phase of the differential dynamic cross-term. The term $N_{prod}(t,f)$ encapsulates the complex noise cross-products resulting from the dual-RX multiplication. The objective of the subsequent processing steps is to isolate the phase contribution originating from the differential dynamic cross-term.

\textbf{1. Normalization ($u[n,k]$):}
The initial operation in the sequential processing of $\tilde{H}(t,f)$ is normalization. This step is performed for each time instant $n$ and subcarrier $k$:
$$u[n,k] = \frac{\tilde{H}[n,k]}{|\tilde{H}[n,k]|} = e^{j\angle \tilde{H}[n,k]}$$
This transformation scales the magnitude of each complex CSI point to unity, thereby isolating its phase component. The phase $\angle \tilde{H}[n,k]$ represents the combined angular information from all signal components and propagated noise. This stage is crucial because Doppler information is inherently encoded in the phase, whereas amplitude fluctuations (due to fading, shadowing, or variations in $|C(t,f)|^2$) are typically not directly related to motion sensing and can introduce confounding effects. By focusing solely on the phase, subsequent processing steps are not influenced by variations in signal strength. A known trade-off of normalization is that in regions of very low signal magnitude (e.g., during deep fades or at low SNR), the noise component, when divided by a small signal magnitude, can be disproportionately amplified, potentially leading to unreliable phase estimates. This step also inherently discards amplitude information, which might be valuable for other sensing modalities such as localization or distinguishing object size.

\textbf{2. Subcarrier Averaging ($\overline{u}[n]$):}
Following normalization, the complex values $u[n,k]$ are averaged across $K$ subcarriers to produce $\overline{u}[n]$:
$$\overline{u}[n] = \frac{1}{K}\sum_{k=0}^{K-1}u[n,k]$$
This process functions as a frequency-domain averaging filter. Its primary purpose is to mitigate the effects of frequency-selective fading, where channel conditions vary significantly across different subcarriers, and to suppress noise components that are uncorrelated across frequency. By coherently combining information from multiple subcarriers, the SNR of the time-varying phase estimate is enhanced. Under the assumption that the propagation channel is approximately flat across the averaged subcarriers (i.e., the maximum delay spread of dominant paths is small compared to the inverse of the subcarrier bandwidth), the time-varying Doppler information is largely preserved. The resulting $\overline{u}[n]$ primarily reflects the average phase evolution across the bandwidth. However, if the flat-fading assumption is violated (e.g., in environments with significant delay spread), this averaging can blur or distort the phase information, potentially losing fine delay-domain features that could be useful for other advanced analyses.

\textbf{3. Unwrap Phase ($\phi_{raw}[n]$):}
The next step involves extracting the phase of the averaged signal $\overline{u}[n]$ and unwrapping it to obtain a continuous phase trajectory:
$$\phi_{raw}[n] = \text{unwrap}(\angle \overline{u}[n])$$
The `angle` function, as typically implemented in programming environments, constrains phase values to a principal interval, such as $(-\pi, \pi]$ radians or $(-180^\circ, 180^\circ]$. However, the true phase changes induced by continuous Doppler shifts accumulate over time, often exceeding this range. The unwrapping operation identifies and rectifies these $2\pi$ (or $360^\circ$) discontinuities by adding or subtracting appropriate multiples of $2\pi$, thereby reconstructing the true, continuous phase progression over time. This continuous trajectory is indispensable for obtaining a valid time derivative in later stages, as discontinuities would lead to erroneous large spikes. The robustness of phase unwrapping is highly dependent on the quality of the input phase signal; even small amounts of residual noise can introduce erroneous $2\pi$ jumps, leading to error propagation that renders subsequent Doppler estimation unreliable.

\textbf{4. Background/Static Subtraction ($\phi[n]$):}
Even after the preceding steps, the unwrapped phase $\phi_{raw}[n]$ typically contains components originating from the static environment (e.g., fixed reflectors, furniture, direct LOS path). These contributions manifest as a fixed phase offset or a very slowly varying baseline. To isolate phase changes exclusively due to dynamic targets, a background phase profile is subtracted:
$$\phi[n] = \phi_{raw}[n] - \phi_{background}[n]$$
The background phase $\phi_{background}[n]$ is usually estimated during a period of known static conditions (e.g., by averaging $\phi_{raw}[n]$ over a static observation window when no motion is present). This operation functions as a high-pass filter in the frequency domain, effectively removing the zero-Doppler component associated with static clutter. The efficacy of this step critically depends on the stationarity of the background environment. If the background changes over time (e.g., due to temperature shifts affecting propagation paths, or subtle, unintended movements) or if the estimation period includes unwanted subtle motions, imperfect subtraction can lead to residual static components that may obscure the dynamic signal. Conversely, an overly aggressive static removal can inadvertently filter out legitimate low-Doppler dynamic signals.

\textbf{5. Noise Filtering ($\phi_{sg}[n]$):}
The phase signal $\phi[n]$, though primarily isolated to dynamic components, invariably retains residual noise and artifacts from previous stages. To prepare for the subsequent differentiation, which inherently amplifies high-frequency components (including noise), a noise filtering step is applied. A Savitzky-Golay (SG) filter is commonly chosen for this purpose, producing $\phi_{sg}[n]$:
$$\phi_{sg}[n] = \mathcal{S}\{\phi[n]\}$$
where $\mathcal{S}$ denotes the SG filtering operation. The SG filter operates by fitting a low-degree polynomial to a sliding window of data points, thereby smoothing the phase trajectory while generally preserving the underlying signal shape and local derivatives better than simpler moving average filters. The selection of filter parameters (window size and polynomial order) is a critical design decision: an overly aggressive filter may smooth out legitimate rapid phase changes associated with high Doppler frequencies, potentially leading to a loss of information or distortion of the true velocity profile. Conversely, insufficient filtering leaves noise that will be amplified by the differentiation process.

\textbf{6. Phase Derivative ($v_{\Delta}[n]$):}
The final stage transforms the clean, filtered dynamic phase into the physically interpretable parameter of differential velocity. The instantaneous differential Doppler frequency, $\Delta\nu(t)$, is fundamentally proportional to the time derivative of the filtered phase. Given that the processed phase $\phi_{sg}(t)$ predominantly represents the phase of the differential dynamic cross-term, its derivative with respect to time yields:
$$\frac{d\phi_{sg}(t)}{dt} \approx 2\pi \Delta\nu(t)$$In a discrete-time system, this derivative is computed using numerical approximation methods (e.g., finite differences). Once $\Delta\nu(t)$ is obtained, the differential velocity $v_{\Delta}(t)$ is directly computed using the known wavelength $\lambda$:$$v_{\Delta}(t) = \frac{\lambda}{2\pi}\Delta\nu(t)$$
This step provides a direct, physically interpretable measure of the target's relative motion. It is important to note that differentiation inherently amplifies any remaining high-frequency noise in the phase signal, making the robustness of all preceding filtering and calibration steps paramount.

\newpage
\section{Experimental methodology}

\subsection{Measurement scenarios}

\begin{figure}[htbp]
    \centering
    \begin{subfigure}[t]{0.45\textwidth}
        \centering
        \includegraphics[width=0.9\textwidth]{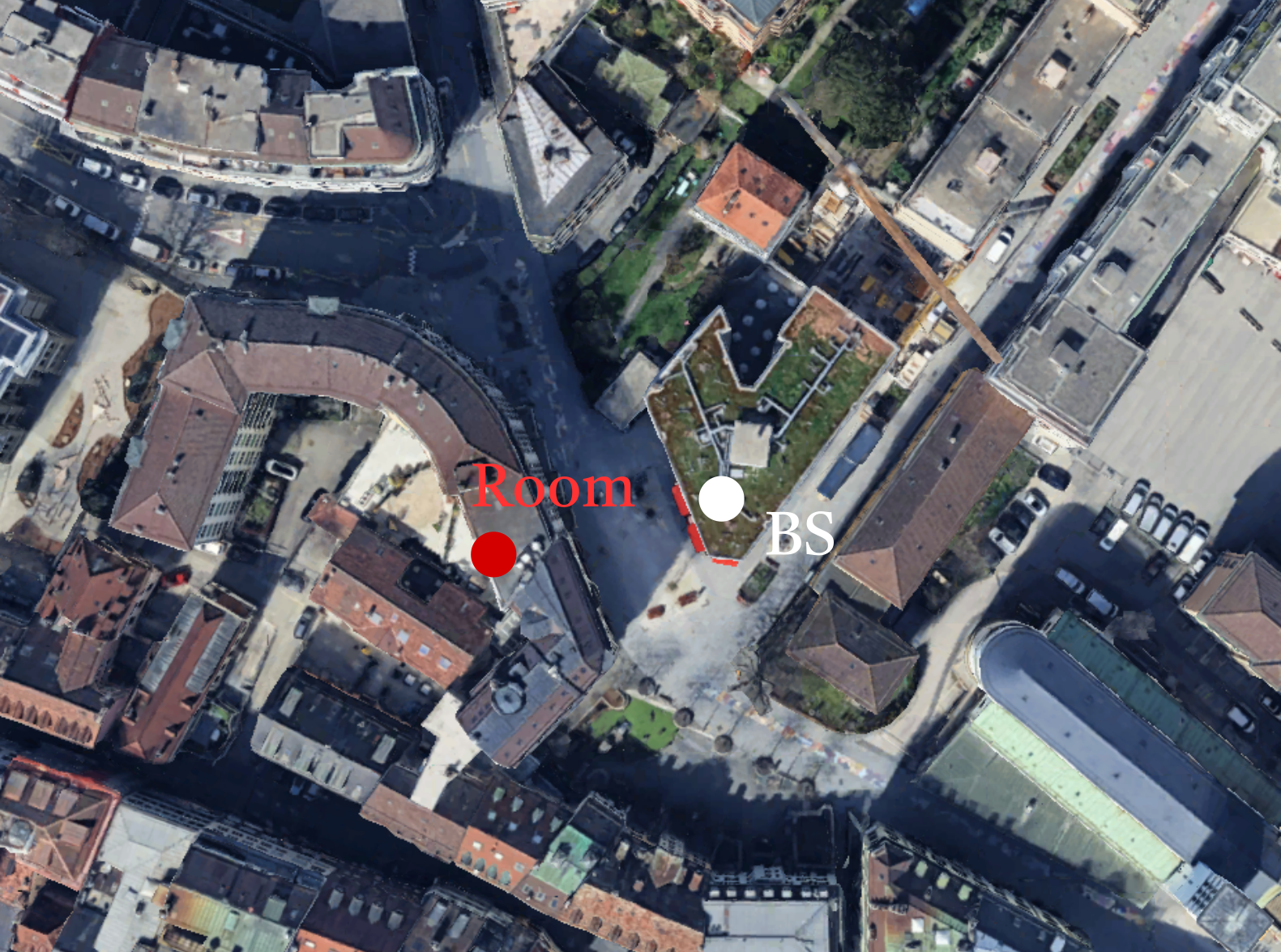}
        \caption{}
        \label{fig:indoor_location}
    \end{subfigure}
    \hfill
    \begin{subfigure}[t]{0.45\textwidth}
        \centering
        \includegraphics[width=0.9\textwidth]{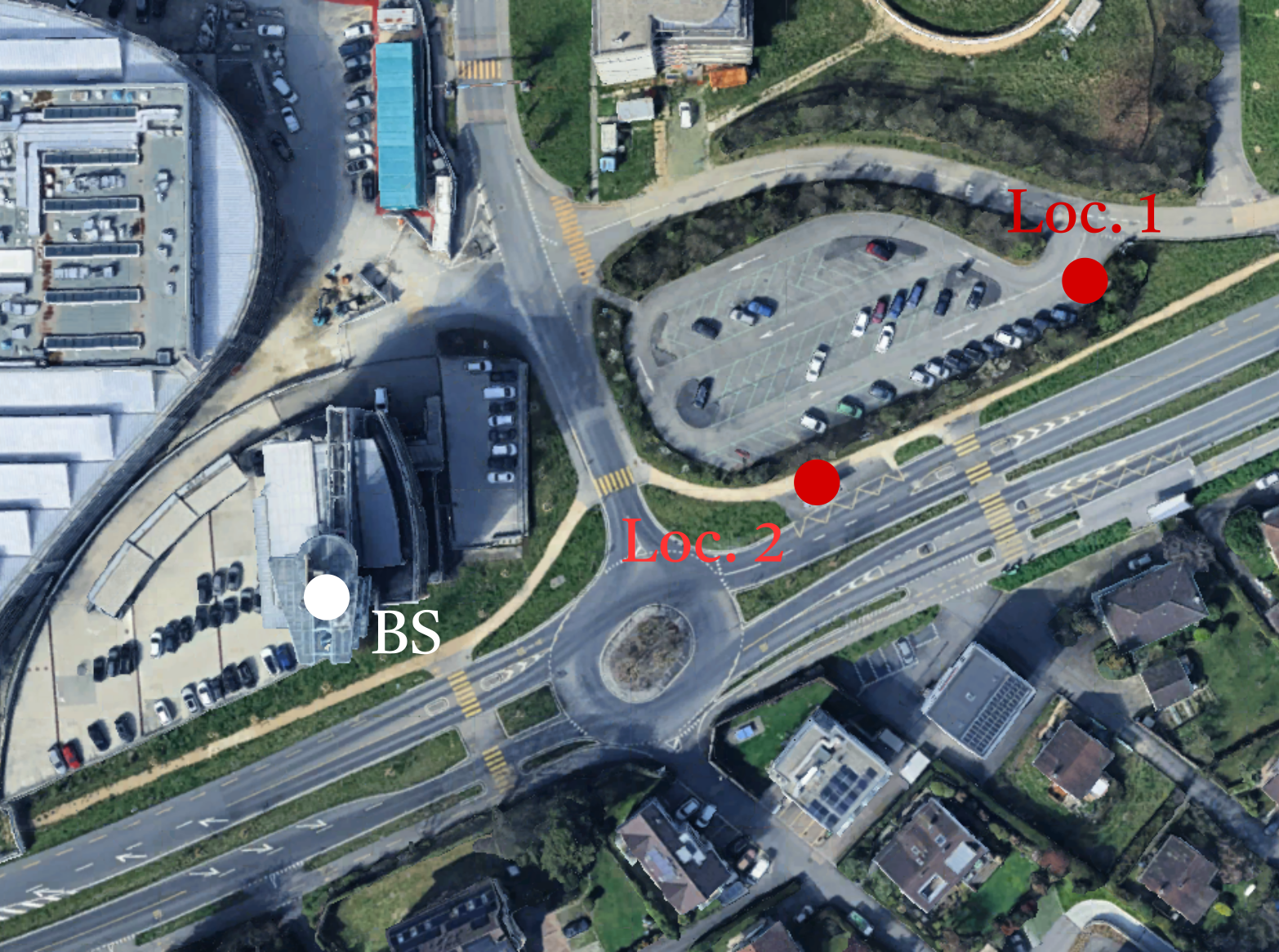}
        \caption{}
        \label{fig:outdoor_location}
    \end{subfigure}
    \caption{Measurement scenarios: (a) indoor and (b) outdoor environments.}
    \label{fig:location_setups}
\end{figure}

The validation of any wireless sensing system necessitates a comprehensive evaluation across diverse environments. This project's experimental methodology involved data acquisition in both controlled indoor settings and varied outdoor scenarios. These include a controlled indoor environment for repeatable measurements, and two distinct outdoor environments: a parking area utilized for pedestrian motion analysis and a roadside setting for vehicular traffic monitoring. These measurements were designed to assess system performance under ideal, repeatable conditions as well as more challenging, real-world complexities. For all measurement scenarios, the LTE BS operated at a downlink frequency of 2.1 GHz, corresponding to EARFCN 203 in LTE Band 1. Satellite imagery for both indoor and outdoor locations, indicating the approximate positions of the BS and receiver antennas, is provided in Figure~\ref{fig:indoor_location} and Figure~\ref{fig:outdoor_location}, respectively.

\subsubsection{Controlled indoor room}

\begin{figure}[htbp]
    \centering
    \begin{subfigure}[t]{0.26\textwidth}
        \centering
        \includegraphics[width=0.92\textwidth]{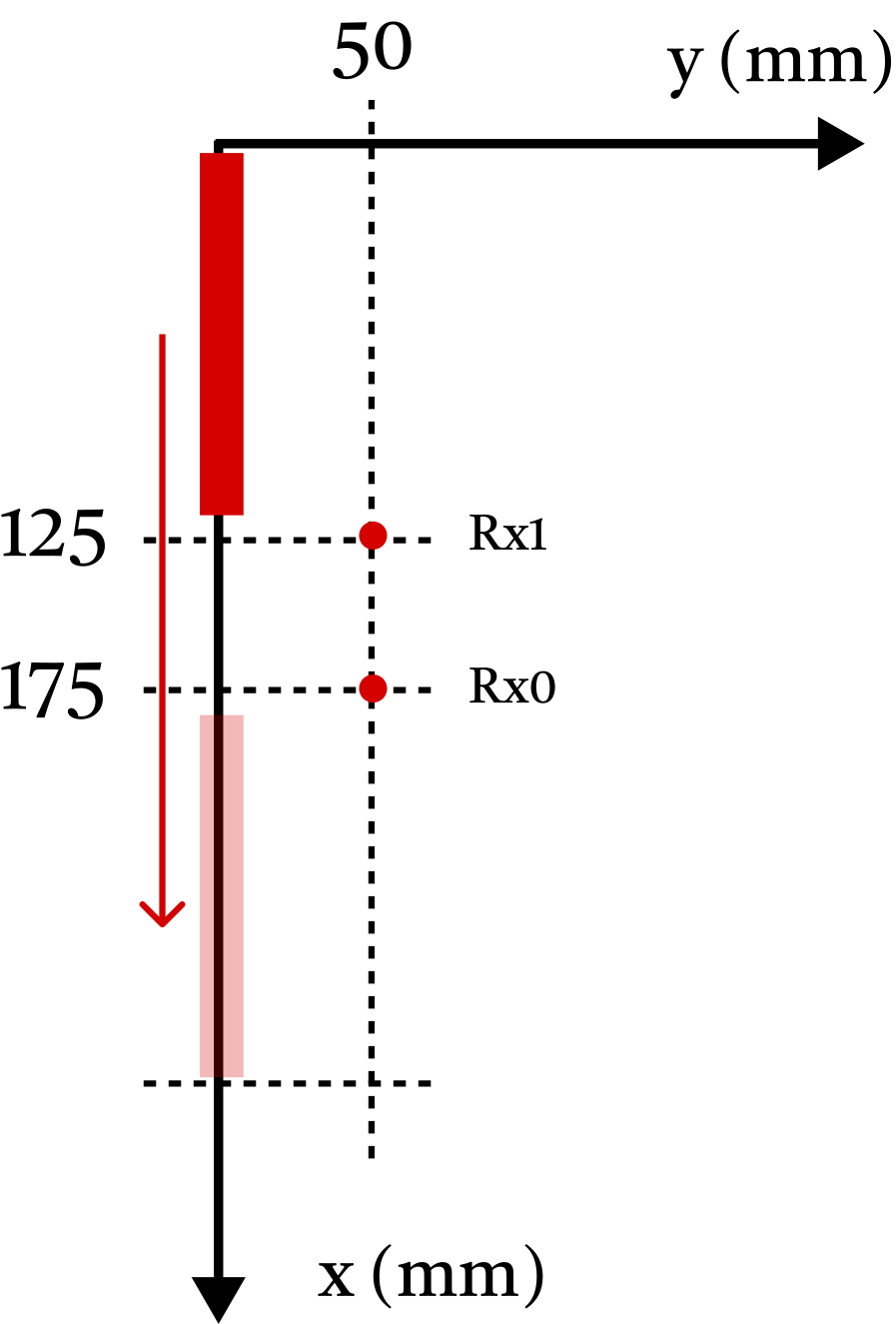}
        \caption{}
        \label{fig:schematic}
    \end{subfigure}
    \hfill
    \begin{subfigure}[t]{0.36\textwidth}
        \centering
        \includegraphics[width=0.92\textwidth]{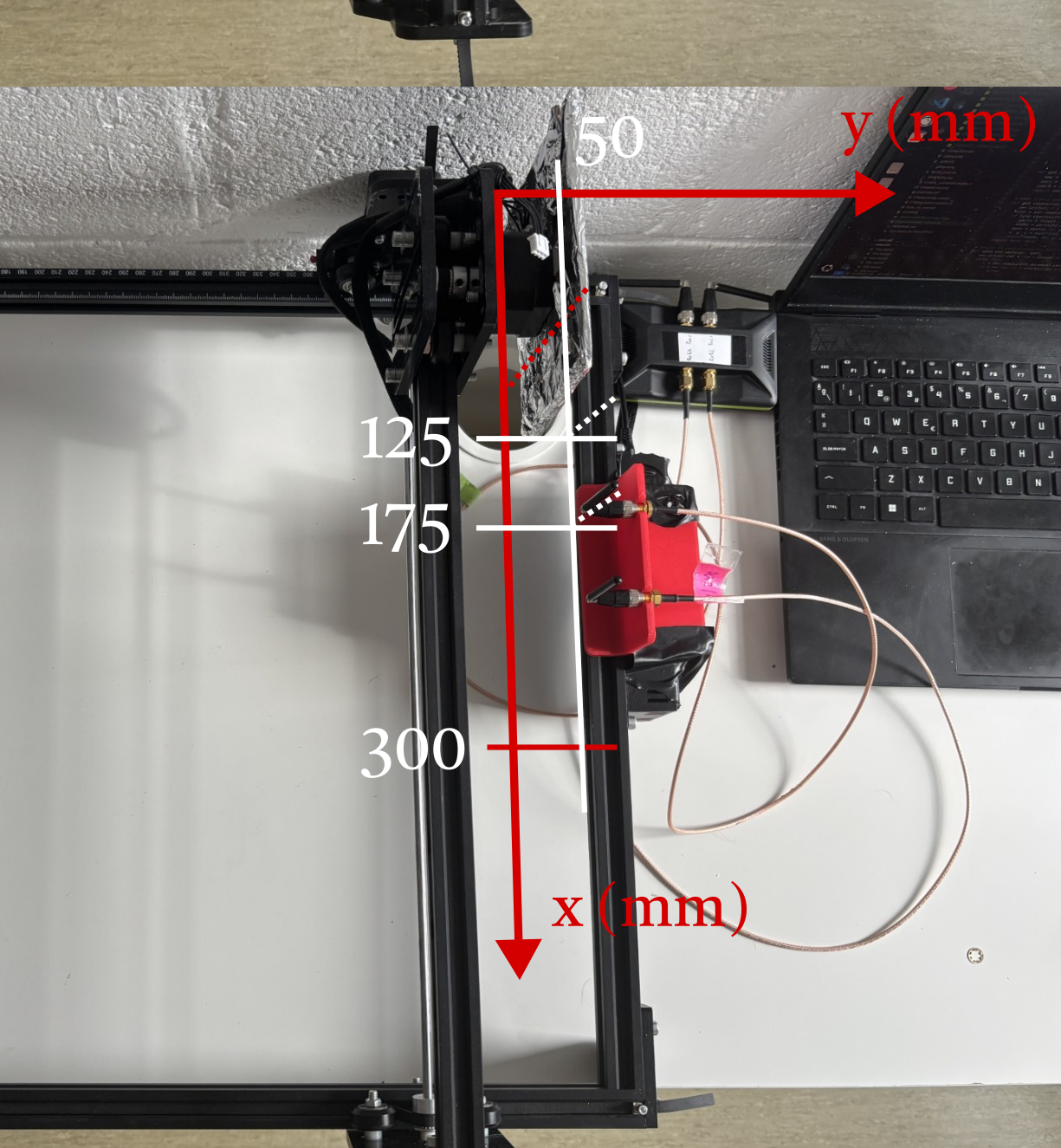}
        \caption{}
        \label{fig:indoor1}
    \end{subfigure}
    \hfill
    \begin{subfigure}[t]{0.36\textwidth}
        \centering
        \includegraphics[width=0.92\textwidth]{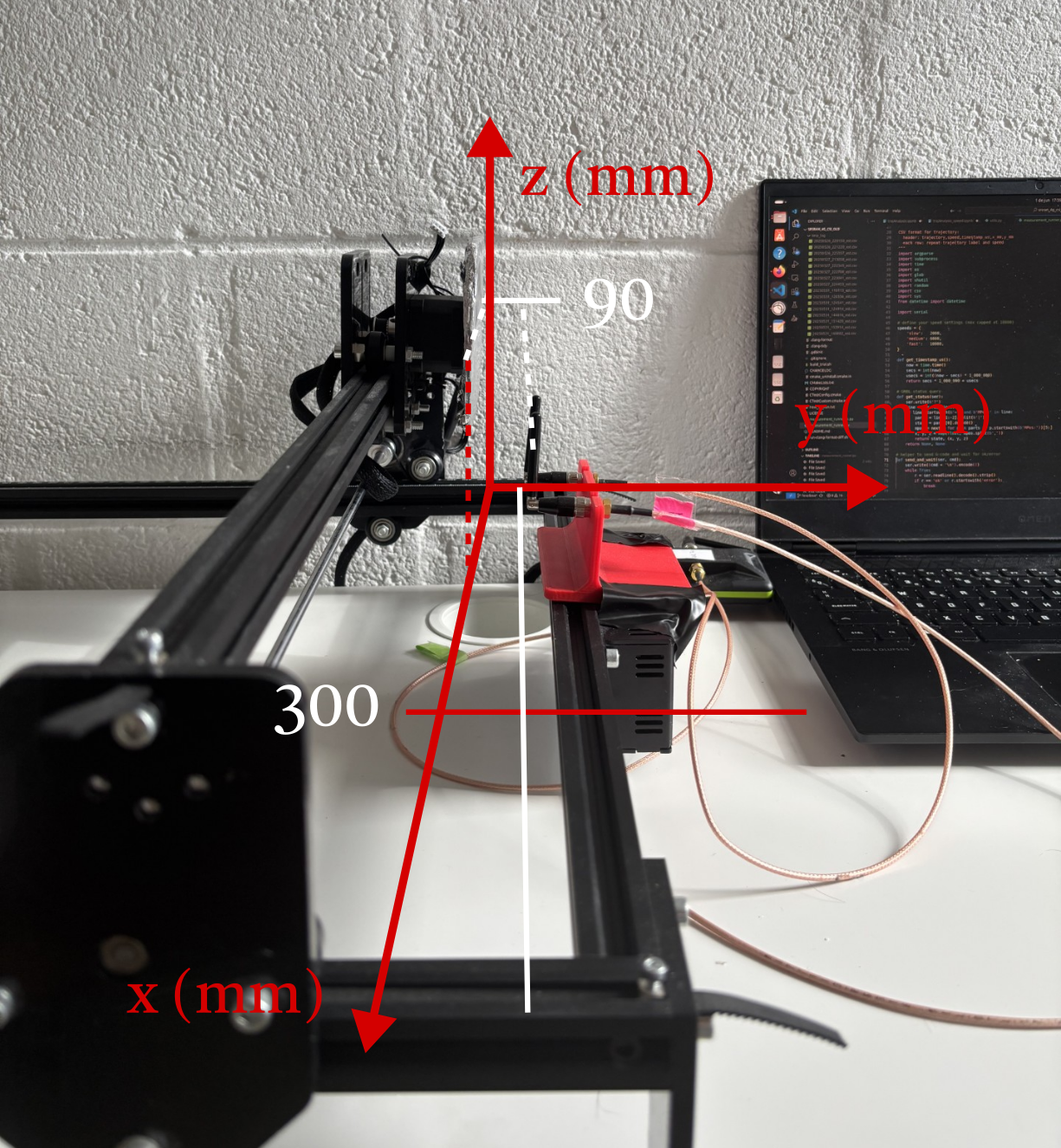}
        \caption{}
        \label{fig:indoor2}
    \end{subfigure}
    \caption{Setup for controlled indoor room scenario. (a) Trajectory schematic. (b) Top view. (c) Side view.}
    \label{fig:indoor}
\end{figure}

The controlled indoor environment was established within a residential room, designed to provide a highly repeatable and isolated setting for initial system validation. As shown in the schematic (Figure~\ref{fig:schematic}) and measurement photographs (Figure~\ref{fig:indoor1}, Figure~\ref{fig:indoor2}), the setup involved a dynamic reflector moving along a precisely defined trajectory. The propagation environment was characterized by Non-Line-of-Sight (NLOS) conditions, with static reflections originating from walls and furniture at distances of 1-3 meters from the measurement area. The dynamic reflector, attached to the linear positioner CNC, operated at a close range, between 5-15 cm from the measurement zone, interacting with the static environment. The primary purpose of this indoor setup was to emulate a vehicle crossing, albeit on a scaled trajectory, to enable accurate and repeatable validation of detection accuracy under isolated conditions.

Measurements in this controlled environment were conducted under various speeds, including 2000, 6000, and 10000 mm/min. The trajectory consisted of a vertical linear motion along the x-axis, where the reflector moved from 0 to 300 mm. The measurements were conducted with the aim of testing the method's capabilities in detection and speed classification. Two distinct types of data capture protocols were implemented:

In the first type, each measurement run was performed with a constant, predefined reflector speed throughout its duration. The second type involved varying the reflector's speed within a single measurement run, allowing for the observation of dynamic changes in the CSI signature under different profiles.

These two measurement protocols were designed to evaluate speed classification performance in both intra-measurement (consistency within the same measurement settings) and inter-measurement (generalizability across different recording sessions) scenarios.

The indoor environment provided a high-quality radio link, characterized by high RSRP and SNR at the receiver location. These stable link conditions facilitated the capture of extensive CSI data over long periods, minimizing environmental interference and ensuring clean, repeatable CSI captures due to the absence of other moving objects.

\subsubsection{Outdoor scenarios}

To evaluate the system's robustness and effectiveness in application-relevant conditions, measurements were extended to two distinct outdoor scenarios: a parking area and a roadside traffic location. These environments were selected to present more complex and less controlled propagation conditions compared to the indoor setup.

\begin{figure}[htbp]
    \centering
    \begin{subfigure}[t]{0.36\textwidth}
        \centering
        \includegraphics[width=0.92\textwidth]{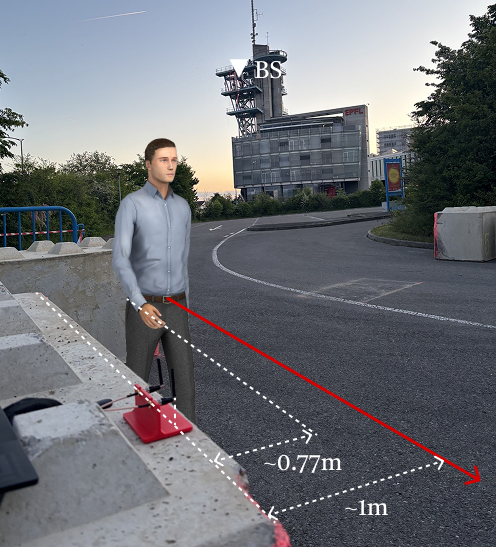}
        \caption{}
        \label{fig:outdoor1}
    \end{subfigure}
    \begin{subfigure}[t]{0.36\textwidth}
        \centering
        \includegraphics[width=0.92\textwidth]{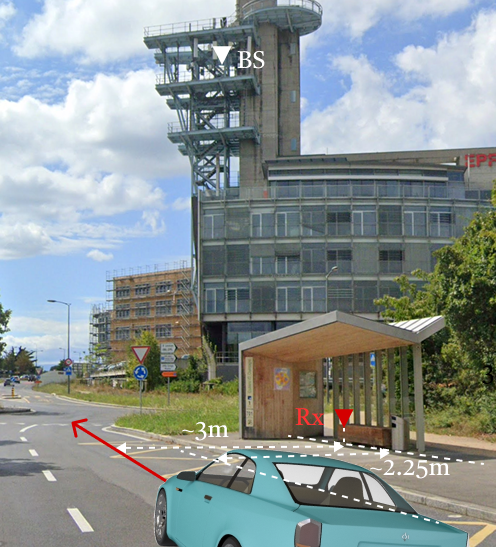}
        \caption{}
        \label{fig:outdoor2}
    \end{subfigure}
    \caption{Setup for outdoor scenarios. (a) Parking area with controlled pedestrian motion. (b) Roadside traffic with real vehicle passes.}
    \label{fig:outdoor}
\end{figure}

The parking area scenario involved controlled pedestrian motion, where human subjects traversed defined linear paths at an approximate distance of 1 meter. In this specific setup, the receiver antennas maintained a LOS path to the moving pedestrians (Figure~\ref{fig:outdoor1}). This configuration enabled the assessment of pedestrian motion detection under relatively clear outdoor conditions and allowed for speed estimation by linking the observed differential Doppler to the known trajectory geometry.

The roadside traffic scenario focused on capturing real vehicle passes along a straight road section located approximately 3 meters from the receiver (Figure~\ref{fig:outdoor2}). This environment inherently presented NLOS conditions, as the direct LOS was obstructed by a bus shelter, adding complexity to signal propagation and Doppler interpretation.

In both outdoor environments, despite the proximity to the BS leading to a high RSRP, the overall SNR was observed to be degraded. This degradation is attributed to the complex multipath propagation and higher levels of RF interference characteristic of urban settings. The compromised SNR in these less controlled environments made it challenging to capture continuous CSI data over extended periods, impacting the volume of data available for proper evaluation of the method under operational conditions.

\subsection{Performance metrics}

The evaluation of the proposed methodology requires a clear set of metrics to quantitatively and qualitatively assess its performance in detecting and characterizing motion. The selection of these metrics is tailored to the specific objectives of traffic monitoring, encompassing event detection, speed classification, and speed estimation.

For assessing event detection performance, two primary metrics are utilized:

\begin{itemize}
    \item \textbf{Detection Rate (DR):} This metric quantifies the proportion of actual events that the system correctly identifies. It is defined as the ratio of true positives to the sum of true positives and false negatives:
    \[
    \text{DR} = \frac{\text{True Positives}}{\text{True Positives} + \text{False Negatives}}
    \]
    This provides insight into the system's sensitivity to genuine events.

    \item \textbf{False Positive Rate (FPR):} This metric indicates the proportion of instances where the system incorrectly reports an event when no actual event occurred. It is calculated as the ratio of false positives to the sum of false positives and true negatives:
    \[
    \text{FPR} = \frac{\text{False Positives}}{\text{False Positives} + \text{True Negatives}}
    \]
    A low FPR signifies the system's reliability and its ability to avoid spurious detections.
\end{itemize}

Speed classification capabilities are evaluated through the use of confusion matrices. These matrices illustrate the distribution of predicted speed categories against their true labels. Importantly, the classification is performed based on relative speed detection, exploiting Doppler dynamics and signal patterns associated with different speed ranges, rather than estimating absolute speeds and assigning them to predefined bins. The resulting classification accuracy reflects the percentage of correctly identified speed categories within these predefined ranges. This evaluation is conducted in two distinct contexts:

\begin{itemize}
    \item \textbf{Intra-measurement classification:} assesses the consistency of speed classification for events within a single recording session, where environmental and processing conditions conditions remain constant.

    \item \textbf{Inter-measurement classification:} evaluates the generalizability of the classification performance across different recording sessions, which may involve varying environmental or processing factors, thereby testing the method's robustness to real-world variability.
\end{itemize}

In scenarios where direct ground truth for error calculation is unavailable (e.g., outdoor measurements), the method's ability to provide reasonable speed estimations is qualitatively noted. The speed of a target is estimated from the peak differential Doppler frequency ($\Delta\nu$) using the relationship derived in Section~3.2:
\[
v_x \approx -\Delta\nu(x_c)\frac{\lambda R_m}{2(x_c - x_0)}
\]
where $v_x$ is the target's velocity component along the x-axis, $\Delta\nu(x_c)$ is the peak differential Doppler at the midpoint of the antenna baseline $x_c$, $\lambda$ is the wavelength, $R_m$ is the distance from the midpoint to each receiver antenna, and $(x_c - x_0)$ is half the antenna separation.

For outdoor measurements, due to the limited number of data samples and the inherent complexity of collecting precise ground truth information in real-world scenarios, the evaluation relies on a combination of visual assessment and direct application of the speed estimation formula. This involves visually checking the consistency of the estimated differential Doppler curves with known time instants of event occurrences (ground truth markings) and validating the estimated speeds against expected ranges. Additionally, the ability of the system to infer the directionality of motion is considered as a qualitative performance aspect.

\section{Results}

\subsection{Controlled indoor room results}

\begin{figure}[htbp]
    \centering
    \includegraphics[width=.7\linewidth]{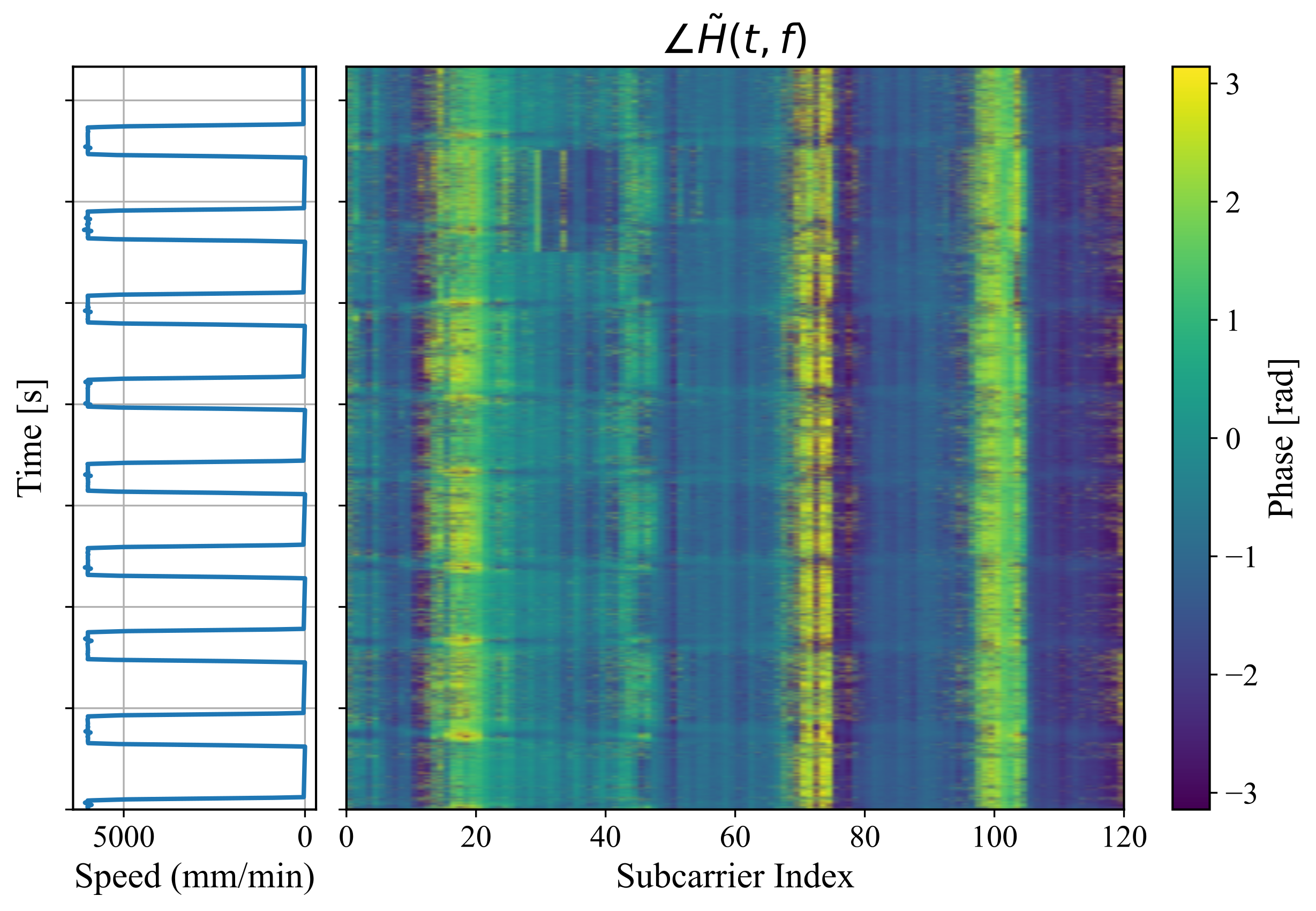}
    \caption{Phase of the composite signal $\tilde{H}(t,f)$ over time, showing phase variations in response to reflector movement at 6000 mm/min.}
    \label{fig:CSI_phase}
\end{figure}

The performance of the proposed methodology was initially evaluated in a controlled indoor environment, providing a baseline for its capabilities in motion detection and speed estimation under isolated conditions. Figure~\ref{fig:CSI_phase} presents the phase of the composite signal $\tilde{H}(t,f)$, which is the result of cross-correlating the signals from both receiver branches. This figure illustrates the phase variations of $\tilde{H}(t,f)$ over time, highlighting how movement (indicated by the reflector speed on the left) induces observable changes in phase. This signal serves as the starting point for the subsequent processing pipeline, which aims to isolate the differential Doppler component from these variations, which also contain static-dynamic components.

\begin{figure}[htbp]
    \centering
    \includegraphics[width=\linewidth]{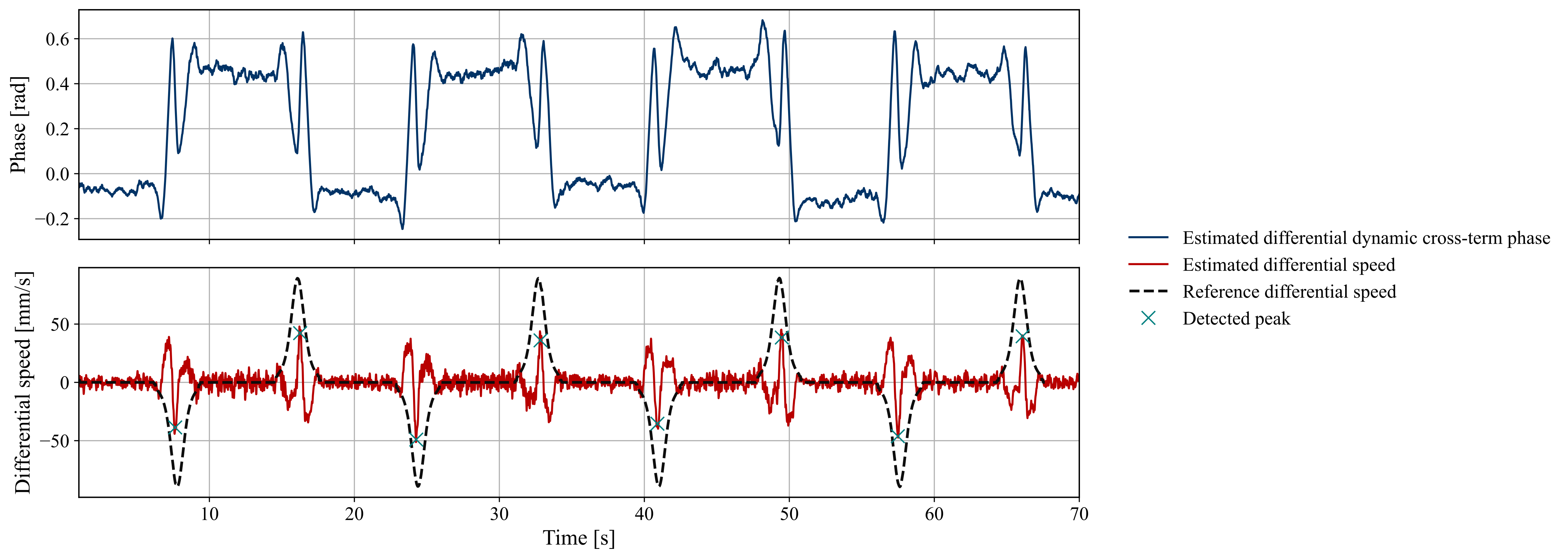}
    \caption{Estimated differential dynamic cross-term phase and corresponding differential speed for reflector movement at 6000 mm/min.}
    \label{fig:curve_example}
\end{figure}

\begin{figure}[htbp]
    \centering
    \includegraphics[width=\linewidth]{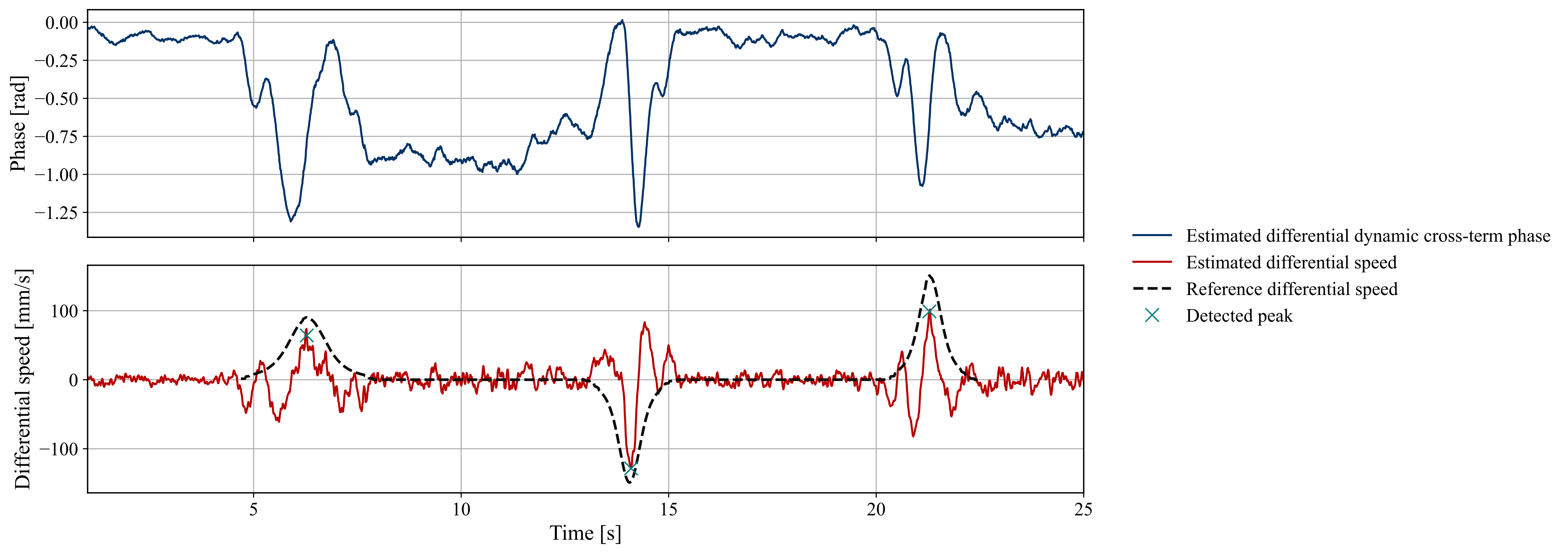}
    \caption{Estimated differential dynamic cross-term phase and corresponding differential speed for reflector movement at different speeds (2000 mm/min and 6000 mm/min).}
    \label{fig:curve_speed_example}
\end{figure}

Figure~\ref{fig:curve_example} presents a representative example of the processed signals for a reflector movement at 6000 mm/min. The top subplot displays the estimated differential dynamic cross-term phase, while the bottom subplot illustrates its corresponding differential speed estimation alongside the ground truth. It is observed that the estimated curves, while closely tracking the general trend of the ground truth, do not perfectly align with the ideal reference. This discrepancy arises because the differential dynamic cross-term, despite the processing pipeline, is not entirely isolated from residual noise and subtle interference. Furthermore, the signal processing chain, particularly filtering operations, inevitably leads to some loss of information, as phase data is highly sensitive to such manipulations. Due to these inherent losses and the challenge of obtaining exact speed values, the evaluation of speed for indoor measurements primarily relies on relative comparisons. The behaviour of the differential speed estimation at varying reflector speeds is illustrated in Figure~\ref{fig:curve_speed_example}. It is evident from this figure that changes in speed result in noticeable variations in peak magnitudes, reinforcing the suitability of using relative speed comparisons rather than absolute values for classification purposes. Nevertheless, both Figure~\ref{fig:curve_example} and Figure~\ref{fig:curve_speed_example} clearly demonstrate that event occurrences and the direction of motion are discernible. The distinct peaks in the differential speed curve provide a robust indicator for event detection, suggesting that a straightforward peak-detection algorithm could effectively identify motion events.

\begin{figure}[htbp]
    \centering
    \includegraphics[width=0.8\linewidth]{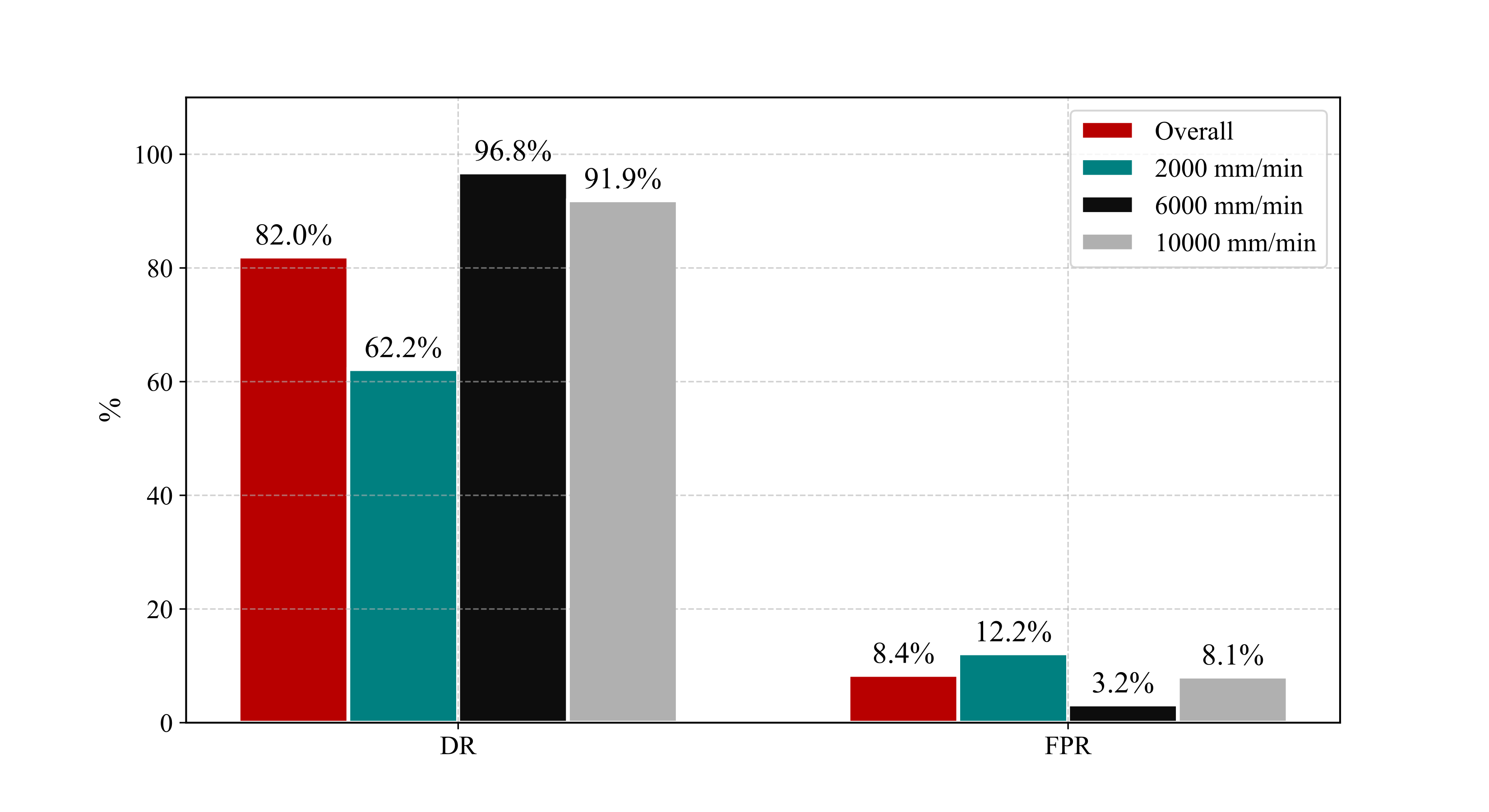}
    \caption{Detection performance across 239 controlled indoor trajectories at different reflector speeds.}
    \label{fig:detection_plot}
\end{figure}

The insights gained from analyzing individual signal traces are further supported by the quantitative detection performance metrics, as summarized in Figure~\ref{fig:detection_plot}. This histogram, based on 239 controlled indoor trajectories, reveals that the DR generally improves with increasing reflector speed. For instance, the overall DR is 82.0\%, but it markedly increases from 62.2\% at 2000 mm/min to 96.8\% at 6000 mm/min and 91.9\% at 10000 mm/min. This trend indicates that at lower speeds, the dynamic echo's amplitude more closely approaches the levels of residual static clutter and additive noise, thereby reducing the confidence in distinguishing genuine motion from background fluctuations. Conversely, for speeds above 6000 mm/min, the calibrated dual-RX CSI pipeline robustly identifies motion in this controlled indoor setting. The FPR also varies with speed, with values of 12.2\% at 2000 mm/min, 3.2\% at 6000 mm/min, and 8.1\% at 10000 mm/min. Notably, for speeds of 6000 mm/min and 10000 mm/min, the sum of DR and FPR approaches 100\%. This suggests that some events classified as false positives might, in fact, be correct detections that are merely slightly offset from the precise ground truth timestamps.

\begin{figure}[H]
    \centering
    \begin{subfigure}[t]{0.49\textwidth}
        \centering
        \includegraphics[width=0.95\textwidth]{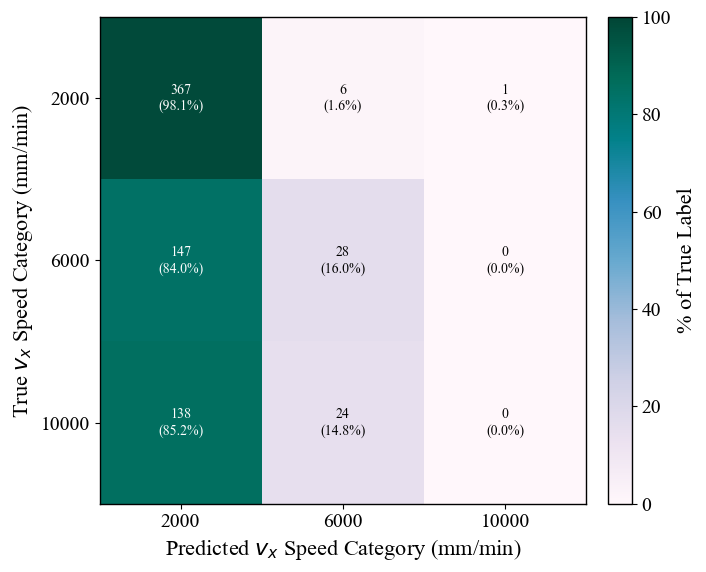}
        \caption{}
        \label{fig:inter_conf_matrix}
    \end{subfigure}
    \hfill
    \begin{subfigure}[t]{0.49\textwidth}
        \centering
        \includegraphics[width=0.95\textwidth]{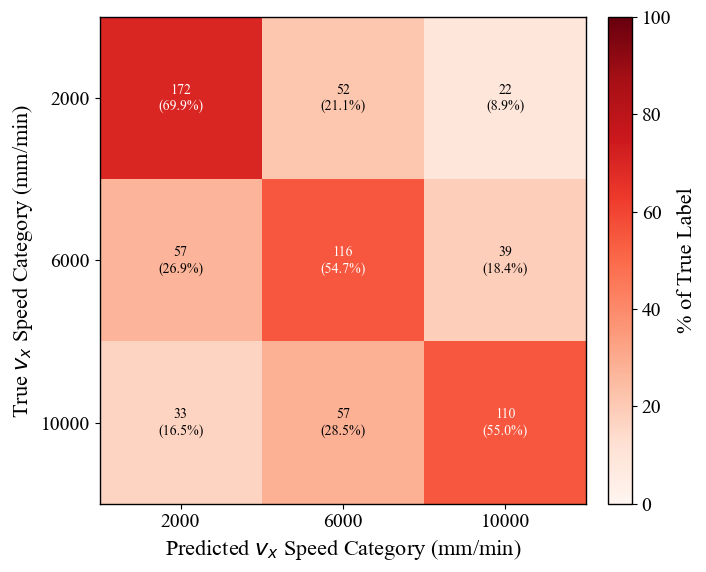}
        \caption{}
        \label{fig:intra_conf_matrix}
    \end{subfigure}
    \caption{Confusion matrices for (a) intra-measurement classification and (b) inter-measurement classification, across 751 controlled indoor trajectories.}
    \label{fig:conf_matrices}
\end{figure}

Beyond detection, the system's ability to classify and estimate speed was also assessed across 751 trajectories. Figure~\ref{fig:conf_matrices} presents confusion matrices for both intra-measurement and inter-measurement speed classification. The intra-measurement classification (Figure~\ref{fig:intra_conf_matrix}) demonstrates high consistency, with a majority of events correctly classified within the same measurement settings. This indicates that once a specific measurement environment is stable, the method can reliably distinguish between different speed categories. However, the inter-measurement classification (Figure~\ref{fig:inter_conf_matrix}) shows a decrease in accuracy. This reduction is attributed to differences between recording sessions, which may necessitate a unique threshold or calibration for each measurement, highlighting challenges in generalization.

\begin{figure}[H]
    \centering
    \includegraphics[width=0.72\linewidth]{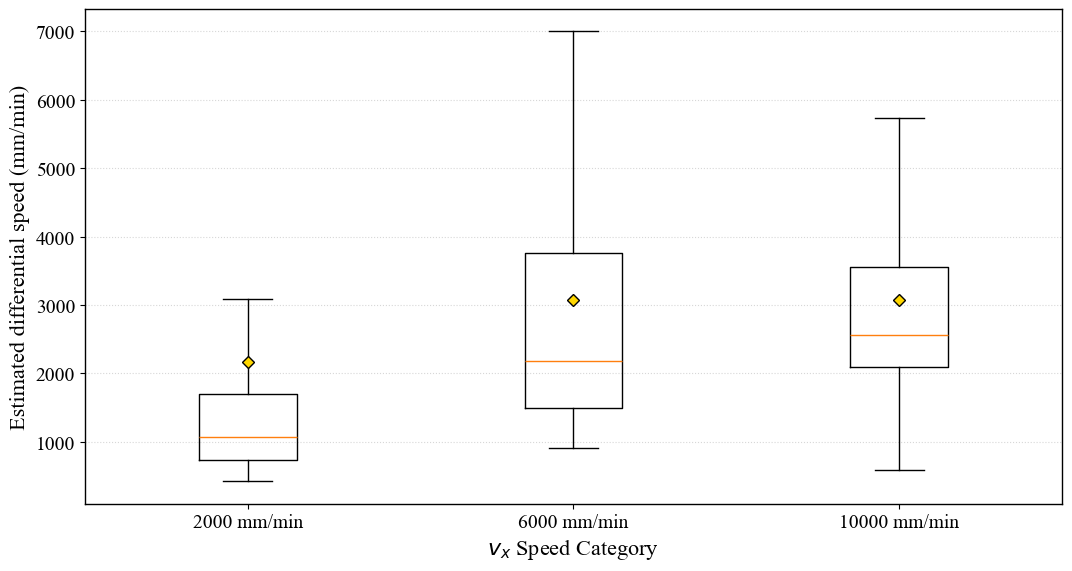}
    \caption{Boxplot of estimated differential speeds for 751 controlled trajectories at different reflector speeds.}
    \label{fig:inter_distribution}
\end{figure}

Further insights into speed estimation are provided by the boxplot in Figure~\ref{fig:inter_distribution}. This plot illustrates the distribution of estimated differential speeds across different true speed categories. While individual speed estimates exhibit variability, the median values within each box (indicated by the orange line and yellow diamond) generally align with the trend of the true speed categories (2000, 6000, and 10000 mm/min). This suggests that despite variations in individual estimates, the method captures the underlying speed trend, providing a relative indication of velocity. The spread of the boxplots (interquartile range) and whiskers indicates the dispersion of estimates, which is larger for higher speeds, reflecting the complexity in precisely estimating faster movements.

\subsection{Parking area \& Roadside traffic evaluation}

Outdoor measurements inherently face challenges such as limited data collection duration and difficulty in obtaining precise ground truth compared to controlled indoor environments. Consequently, the evaluation methodology emphasizes qualitative assessments of processed phase curves, complemented by quantitative speed estimations based on the relationship derived in Section 3.2, adapted here for using differential speed ($v_{\Delta}$):

\begin{equation}
|v_x| \approx |v_{\Delta}(x_c)| \frac{R_m}{|2(x_c - x_0)|}
\end{equation}

\begin{figure}[htbp]
 \centering
\includegraphics[width=0.8\linewidth]{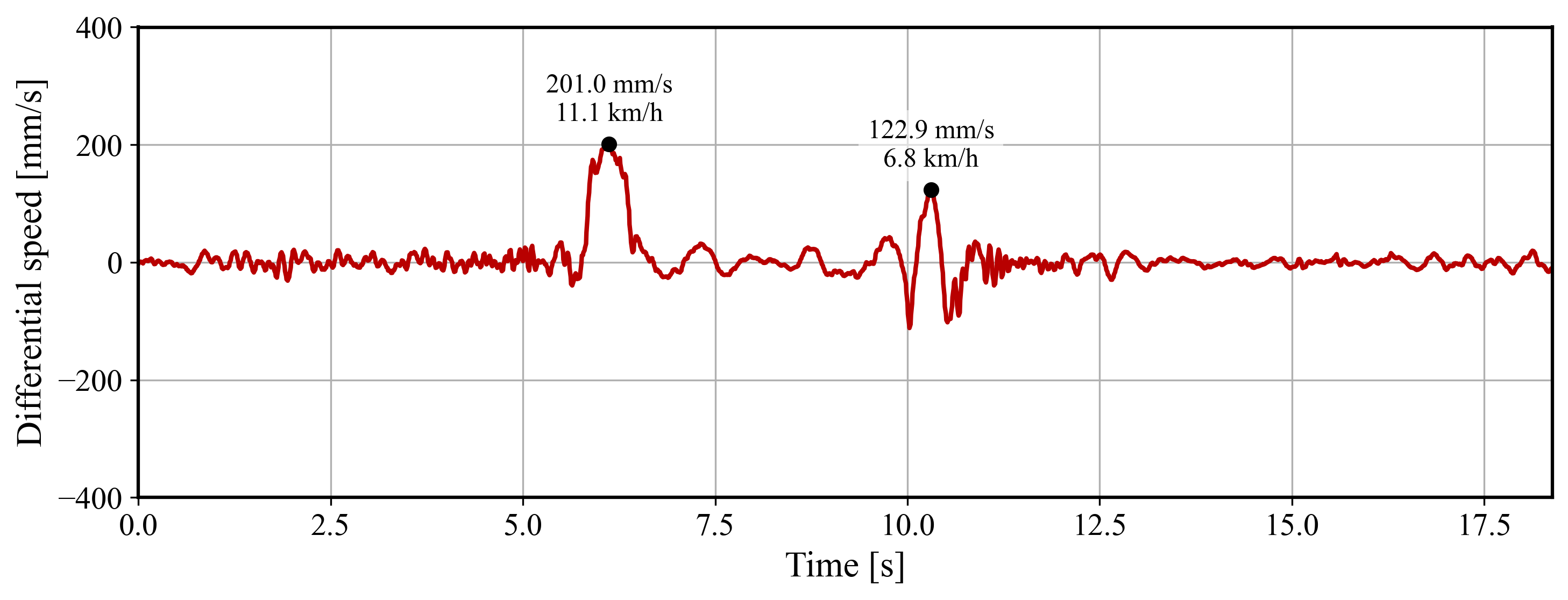}
\caption{Parking area pedestrian trajectory estimated differential speed evaluation.}
\label{fig:parking_pedestrian}
\end{figure}

Figure~\ref{fig:parking_pedestrian} illustrates the estimated differential speed for a pedestrian trajectory in the parking area, involving a back-and-forth traversal. Analysis reveals inconsistency in motion direction detection, as evidenced by non-opposing signs in estimated differential speed values for the two trajectory segments. Despite this limitation, the estimated speeds remain within a reasonable range for pedestrian motion. With parameters $R_m = 0.77~\mathrm{m}$ and $|2(x_c - x_0)| = 0.050~\mathrm{m}$, calculated pedestrian speeds are:

\begin{align*}
v_{\Delta} &= 201~\mathrm{mm/s}: \quad |v_x| \approx 0.201 \cdot \frac{0.77}{0.050} = 3.095~\mathrm{m/s} \approx 11.14~\mathrm{km/h} \\
v_{\Delta} &= 122.9~\mathrm{mm/s}: \quad |v_x| \approx 0.1229 \cdot \frac{0.77}{0.050} = 1.893~\mathrm{m/s} \approx 6.81~\mathrm{km/h}
\end{align*}

These estimations align well with typical pedestrian velocities.

\begin{figure}[htbp]
 \centering
\includegraphics[width=0.8\linewidth]{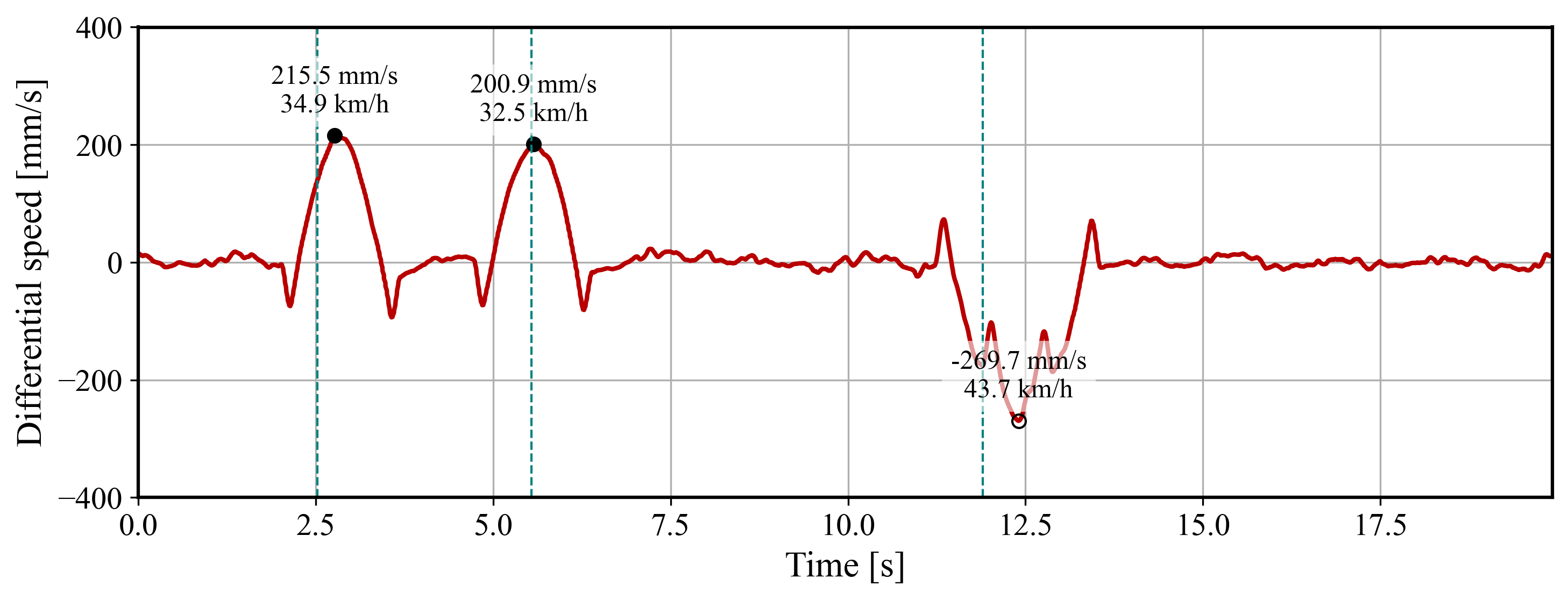}
\caption{Roadside traffic vehicle trajectory estimated differential speed evaluation.}
\label{fig:roadside_vehicle}
\end{figure}

Figure~\ref{fig:roadside_vehicle} presents results from roadside vehicle trajectory analysis. Explicit markers corresponding to vehicle passage events align closely with peaks in the estimated differential speed curve, demonstrating effective event detection. Using $R_m = 2.25~\mathrm{m}$ and $|2(x_c - x_0)| = 0.050~\mathrm{m}$, vehicle speeds are calculated as:

\begin{align*}
v_{\Delta} &= 215.5~\mathrm{mm/s}: \quad |v_x| \approx 0.2155 \cdot \frac{2.25}{0.050} = 9.698~\mathrm{m/s} \approx 34.91~\mathrm{km/h} \\
v_{\Delta} &= 200.9~\mathrm{mm/s}: \quad |v_x| \approx 0.2009 \cdot \frac{2.25}{0.050} = 9.041~\mathrm{m/s} \approx 32.55~\mathrm{km/h} \\
v_{\Delta} &= 269.7~\mathrm{mm/s}: \quad |v_x| \approx 0.2697 \cdot \frac{2.25}{0.050} = 12.137~\mathrm{m/s} \approx 43.69~\mathrm{km/h}
\end{align*}

These values are within expected vehicle speed ranges. Nevertheless, an inconsistency arises in the directional sign of the third detected peak, indicating a potential error either in the direction detection or speed estimation. The current calculation assumes a constant $R_m = 2.25~\mathrm{m}$, appropriate for the nearest traffic lane. If the actual lane distance were larger for the third event, the resulting speed estimation would significantly increase, suggesting the need for more precise distance characterization for accurate speed estimations in multilane traffic scenarios.

Overall, the outdoor evaluations underscore several key findings. Despite the inherent noise and complexity associated with urban environments, phase variations due to target movements remain distinctly observable. Reliable speed estimations are achievable even without explicit ground truth measurements for direct error calculations. Nonetheless, the methodology demonstrates limitations in accurately determining motion direction, particularly in complex or reciprocal trajectory scenarios, warranting further refinement of direction inference techniques.

\newpage
\section{Discussion}

\subsection{Impact of slow motion and multipath fading}

The performance of passive wireless sensing methodologies, particularly those reliant on Doppler shifts, is significantly influenced by challenging channel conditions such as the presence of slow-moving targets and pervasive multipath fading. These factors introduce complexities that can compromise the accuracy of detection and estimation.

Regarding slow motion, the theoretical channel model for $\tilde{H}(t,f)$, derived after dual-receiver multiplication, highlights the potential for interference from static channel components. Specifically, terms such as the static-static product ($H_{s,1}H_{s,0}^{*}$) and static-dynamic cross-products ($H_{s,1}a_{0}^{*}e^{j\Phi_{sd0}(t,f)}$ and $H_{s,0}^{*}a_{1}e^{j\Phi_{sd1}(t,f)}$) can overlap with the desired differential dynamic cross-term. When a target moves at very low speeds, the magnitude of its induced Doppler shift is small, resulting in a dynamic echo amplitude that closely approaches the levels of background static clutter and additive noise. This proximity in signal strength makes it challenging to distinguish the subtle motion signature from the stationary environment. Empirical evidence from the controlled indoor measurements supports this: the DR for the slowest tested speed, 2000 mm/min, was notably lower at 62.2\% compared to higher speeds, as observed in Figure~\ref{fig:detection_plot}. This indicates that the method's ability to discern motion from background noise is compromised at very low velocities.

Multipath fading, a prevalent characteristic of complex environments, particularly in urban outdoor settings, introduces additional complexities. In such scenarios, the signal propagates via multiple paths, resulting in numerous delayed and attenuated replicas arriving at the receiver. This can lead to the constructive and destructive interference of signals, causing rapid fluctuations in CSI amplitude and phase. While the current channel model explicitly focuses on a single dominant dynamic path, realistic multipath conditions may involve several distinct reflections from a single moving target or from other dynamic elements. These multiple dynamic paths can have slightly different delays and Doppler shifts, complicating the isolation of a single, clean differential Doppler signature. Consequently, multipath fading can compromise the accuracy of speed estimation and the reliability of direction detection, as evidenced by the observed inconsistencies in pedestrian directionality in outdoor scenarios. Beyond its impact on signal characteristics, severe multipath fading, combined with general RF interference in urban environments, led to a degraded SNR during outdoor measurements. This compromised SNR made it difficult for the receiver to maintain stable synchronization with the base station and capture continuous, high-quality CSI data over extended periods, thereby impacting the quantity and reliability of collected measurement data.

\subsection{Scalability and cost}

The pervasive deployment of existing cellular network infrastructure offers a compelling foundation for scaling passive traffic monitoring systems, providing an alternative to traditional, often infrastructure-intensive, solutions. The proposed methodology, leveraging ambient LTE signals and a low-cost receiver architecture, inherently possesses attributes conducive to wide-area deployment.

The system's scalability primarily stems from its passive nature; it operates by observing signals already broadcast by cellular BS, eliminating the need for dedicated transmitters. This allows for the deployment of numerous passive receivers across a target area, each monitoring traffic without active transmission, which is advantageous for covertness and spectrum efficiency. Such a distributed network of receivers could provide comprehensive coverage over extensive road segments or urban areas by strategically placing measurement units along existing cellular footprints. This contrasts with conventional methods like inductive loops, which require intrusive road excavation for each detection point, or dedicated radar systems, which involve localized active emissions. However, scaling such a system introduces several complexities. Ensuring precise time synchronization across multiple geographically separated passive receivers is challenging, as relative phase errors directly impact Doppler estimation accuracy in a distributed sensing network. Furthermore, the passive nature means the system has no control over the cellular network's behavior. Interference management from dense cellular traffic or dynamic network reconfigurations (e.g., beamforming changes, power control) can compromise the quality and consistency of the ambient signals used for sensing. 

Regarding cost-effectiveness, the current system utilizes a SDR platform (LimeSDR USB with LMS7002M) primarily for research and development flexibility. In a real-world, large-scale application, a dedicated application-specific integrated circuit (ASIC) would replace the SDR. The LMS7002M, a field-programmable RF (FPRF) chip, represents a highly flexible solution, but its mass-production cost as a fixed-function ASIC performing only the necessary CSI acquisition and initial processing would be considerably lower. Such a specialized chip would leverage economies of scale similar to those seen in mass-produced cellular modem chips, which are significantly more affordable than development-oriented SDRs. This leads to a lower per-unit cost for the receiver hardware and potentially significantly lower initial deployment costs, as it avoids the substantial civil engineering work associated with inductive loops.

Comparing the potential cost of such a dedicated CSI sensing chip to other consolidated radar chips, several factors emerge. While the cost of automotive radar chips (e.g., mmWave FMCW radar) has decreased with volume production, they still represent a higher unit cost compared to what a passive cellular CSI receiver could achieve. This is primarily because radar chips perform active transmission and dedicated signal processing for range, velocity, and angle estimation, often requiring specialized antennas and higher power budgets. In contrast, a passive CSI receiver performs only reception and specific baseband processing. Despite these hardware advantages, the development of robust and adaptive algorithms for CSI-based sensing that can operate reliably across highly variable and uncontrolled cellular environments requires a substantial initial investment in research and algorithm development. This upfront cost in software and signal processing expertise must be weighed against the long-term benefits of reduced infrastructure dependency and potentially lower per-point deployment costs.

\newpage
\section{Future work}

\subsection{Extended outdoor campaigns with ground-truth}

A critical next step involves conducting extended measurement campaigns in diverse outdoor propagation environments. The current outdoor evaluations, while demonstrating feasibility, were limited by data collection duration and the absence of comprehensive ground truth. Future work should prioritize prolonged data acquisition in various real-world scenarios, including different urban densities, roadside configurations, and even weather conditions. Crucially, these campaigns must incorporate ground-truth validation mechanisms, such as synchronized video recordings or established traffic sensors, to enable precise quantitative comparison of estimated speeds and detection events against actual values. This complete evaluation is essential to assess the method's generalization capabilities across a wider spectrum of environmental complexities, moving beyond the controlled conditions of the indoor room. Ultimately, these extended campaigns are fundamental for evaluating the system's potential for real-time live traffic monitoring, where continuous and accurate performance is paramount.

\subsection{AoA \& AI methodology enhancement}

To advance the system's analytical capabilities beyond basic detection and speed estimation, integrating Angle-of-Arrival (AoA) analysis represents a significant enhancement. By estimating the direction from which reflected signals arrive, AoA techniques can enable more sophisticated functionalities such as path tracking of individual vehicles and discrimination between multiple concurrent moving objects within the sensing zone. This would provide a richer understanding of traffic flow dynamics, moving beyond aggregate statistics. Concurrently, exploring Machine Learning (ML) methodologies offers a powerful approach to further refine the detection and classification algorithms. The observed CSI patterns, particularly those resulting from static-dynamic combinations, are not always perfectly clean or easily interpretable by purely information-theoretic models. ML algorithms, such as deep neural networks, possess the capacity to learn complex, non-linear relationships within these noisy and composite CSI patterns. This could lead to improved robustness against environmental variability and enhanced accuracy in classifying traffic conditions, especially where analytical models struggle with the inherent complexities of real-world signal interactions.

\subsection{Real-time embedded implementation and edge deployment}

The transition from a research prototype to a practical traffic monitoring solution necessitates a focus on real-time embedded implementation and efficient edge deployment. The current processing pipeline, while effective, requires computational resources that may not be feasible for widespread, low-cost distributed sensing nodes. Future work would concentrate on optimizing the signal processing algorithms for execution on resource-constrained embedded platforms. This involves exploring efficient numerical methods, fixed-point arithmetic implementations, and hardware acceleration techniques. Deploying these optimized receivers at the network edge, closer to the data source, would minimize backhaul requirements and reduce latency, enabling genuine live traffic monitoring. Challenges such as power consumption management for continuous operation, robust data transmission from edge nodes, and secure, remote management of distributed sensing units must be addressed to ensure the viability and scalability of a real-time, passive cellular traffic monitoring system.

\newpage
\section{Conclusion}

This project embarked on the development and evaluation of a passive sensing system for traffic monitoring, leveraging the extense infrastructure of LTE cellular networks. Driven by the motivation to provide a non-intrusive, scalable, and cost-efficient alternative to conventional traffic surveillance methods, the core objective was to accurately detect and classify vehicular motion by analyzing CSI from ambient LTE signals. The methodology aligned with the emerging paradigm of ISAC, seeking to integrate sensing capabilities into existing communication frameworks.

The theoretical foundation of this work involved a detailed analysis of the wireless channel in dynamic environments. The limitations of single-receiver CSI systems, primarily the corruption of motion-induced phase information by hardware impairments ($C(t,f)$), were rigorously established. This led to the adoption of a dual-RX channel model, where the multiplication of CSI from two closely spaced receivers effectively cancels common hardware-induced phase noise, yielding a composite signal ($\tilde{H}(t,f)$). Crucially, this composite signal contains a differential dynamic cross-term whose phase derivative is directly proportional to the differential Doppler frequency ($\Delta\nu$), an indicator of target velocity. Mathematical derivations demonstrated that for a target traversing a path perpendicular to the receiver baseline, the differential Doppler exhibits a distinct peak at the midpoint, providing a robust signature for event detection and speed estimation. The predominance of this differential dynamic term over static and static-dynamic cross-terms was shown to rely on effective background subtraction and the relative strength of dynamic reflections.

The system's architecture comprised a flexible hardware platform and a data analysis methodology. The LimeSDR USB with its LMS7002M RF transceiver served as the core SDR, providing the necessary dual-channel capability and frequency support for CSI acquisition. The srsRAN 4G software facilitated the emulation of an LTE UE, enabling the capture of fine-grained, complex-valued CSI matrices. The subsequent signal processing pipeline for differential Doppler extraction involved a sequence of stages: normalization, subcarrier averaging, phase unwrapping, background subtraction, noise filtering, and finally, phase differentiation. Each step was justified by its role in progressively isolating the desired motion signature from noise and clutter, transforming raw CSI into a quantifiable differential velocity.

Experimental validation was conducted across controlled indoor and diverse outdoor scenarios. Indoor tests, utilizing a linear positioner to simulate scaled vehicle movements, demonstrated high DR (over 90\%) for medium to high speeds (6000 mm/min and 10000 mm/min), confirming the system's ability to identify motion in controlled settings. However, detection performance was observed to be compromised at very low speeds (2000 mm/min), where dynamic echo amplitudes closely matched residual static clutter, highlighting a challenge in distinguishing subtle motion. Speed classification, evaluated through confusion matrices, showed high consistency within single measurement sessions (intra-measurement) but a reduction in accuracy across different sessions (inter-measurement), indicating variability that affects generalization. Outdoor evaluations, conducted in a parking area for pedestrian motion and a roadside for vehicular traffic, revealed that despite noisier and more complex environments, phase variations attributable to moving targets remained observable. Reasonable speed estimations were obtained for both pedestrians (e.g., 6.81 km/h to 11.14 km/h) and vehicles (e.g., 32.55 km/h to 43.69 km/h), even without precise ground truth for error calculation. A key limitation identified was the inconsistent detection of motion direction, particularly for back-and-forth pedestrian movements, and potential inaccuracies in speed estimation for vehicles in complex multi-lane scenarios.

The discussion of these results highlighted the inherent trade-offs. The system's sensitivity to slow motion is constrained by the signal-to-clutter ratio, where low-Doppler shifts can be masked by persistent static echoes. Multipath fading in outdoor environments not only complicates the isolation of a single dynamic path but also degrades the overall SNR, making robust CSI data acquisition challenging. Despite these technical hurdles, the passive nature of the system presents significant advantages in terms of scalability and cost-effectiveness compared to traditional traffic monitoring solutions. Its reliance on existing LTE infrastructure eliminates the need for dedicated transmitters and intrusive road installations, potentially leading to lower per-unit hardware costs in a mass-produced ASIC form factor. Furthermore, the development of robust algorithms capable of operating reliably across highly variable and uncontrolled cellular environments requires a substantial initial investment in research and algorithm development.

In conclusion, this project successfully demonstrated the feasibility of passive traffic monitoring using LTE CSI, validating the theoretical models and the signal processing pipeline in both controlled and real-world conditions. While the system shows promise for event detection and speed estimation, particularly at higher velocities, it also underscores critical areas for future work. These include conducting extended outdoor measurement campaigns with comprehensive ground-truth validation to rigorously assess robustness in diverse environments, enhancing the methodology with AoA and ML techniques for path tracking and improved classification, and focusing on real-time embedded implementation and edge deployment for practical, scalable traffic monitoring applications. By addressing these challenges, the vision of ubiquitous, low-cost, and non-intrusive traffic monitoring through cellular sensing can be further realized.

\newpage
\printbibliography

%
%

\end{document}